\documentclass[final,1p,times]{elsarticle}

\usepackage{lineno}
\usepackage{graphicx}
\usepackage{mathtools}
\usepackage{natbib}
\usepackage{xstring}
\usepackage{placeins}
\usepackage{amssymb}
\usepackage{amsmath}
\usepackage{xcolor}
\usepackage{xcolor}

\usepackage{color}      
\usepackage[colorlinks=true,
            linkcolor=BrickRed,
            urlcolor=blue,
            citecolor=blue]{hyperref}
            
\usepackage{multicol}
\usepackage{lipsum}
\usepackage{amsthm}

\usepackage[utf8]{inputenc}
\usepackage[T1]{fontenc}
\usepackage{amsmath,amssymb,amsthm}
\usepackage{mathtools}
\usepackage{bm}
\usepackage{tikz}
\usepackage{pgfplots}
\usepackage{booktabs}
\usepackage{array}
\usepackage{geometry}
\usepackage{hyperref}
\usepackage{cleveref}
\usepackage{xcolor}
\usepackage{float}
\usepackage{enumitem}
\usepackage{caption}
\usepackage{subcaption}
\usepackage{graphicx}
\graphicspath{{figures/}}
\usepackage{placeins}
\usepackage{morefloats}  
\usetikzlibrary{shapes.geometric, arrows.meta, positioning, calc, fit, backgrounds, decorations.pathreplacing, patterns, matrix}

\usepackage{algorithm}
\usepackage{algpseudocode}

\definecolor{bayesblue}{RGB}{70,130,180}
\definecolor{detgreen}{RGB}{60,150,60}
\definecolor{basisorange}{RGB}{230,140,50}
\definecolor{outputred}{RGB}{180,60,60}
\definecolor{lightblue}{RGB}{220,235,250}
\definecolor{lightgreen}{RGB}{220,245,220}
\definecolor{lightorange}{RGB}{255,240,220}
\definecolor{lightred}{RGB}{255,230,230}

\newcommand{\E}{\mathbb{E}}

\newcommand{\KL}{D_{\mathrm{KL}}}

\usepackage{pgfplots}
\pgfplotsset{compat=1.18}
\newcommand{\Shat}{\widehat{\mathbf{S}}}
\newcommand{\What}{\widehat{\mathbf{W}}}

\biboptions{sort&compress}

\journal{}

\begin{document}

\begin{frontmatter}

\title{Bayesian Kolmogorov--Arnold networks for trustworthy Reynolds-stress closures}
\author[mymainaddress]{Ali Eidi\corref{cor}}
\author[mymainaddress]{Kherlen Jigjid}
\author[mymainaddress]{Tyler Buchanan}
\author[mymainaddress]{Richard P. Dwight\corref{cor}}


\cortext[cor]{Corresponding authors: Ali Eidi (a.eidi@tudelft.nl) and Richard P. Dwight (r.p.dwight@tudelft.nl).}

\address[mymainaddress]{Department of Flow Physics and Technology, Faculty of Aerospace Engineering, Delft University of Technology, 2629 HS Delft, Netherlands}

\begin{abstract}

Data-driven turbulence closures can improve Reynolds-averaged Navier--Stokes
(RANS) predictions, but their practical use is limited by two related issues:
the learned constitutive relation is often difficult to inspect, and its
predictions are typically provided without a reliable measure of uncertainty.
We develop a Bayesian tensor-basis Kolmogorov--Arnold network (BKAN) for
Reynolds-stress anisotropy that addresses both aspects within a single closure
formulation. Learnable univariate spline functions on the network edges provide
an inspectable representation that can, where sufficiently simple, be
approximated symbolically. Variational posteriors over the spline coefficients
provide a predictive distribution that separates epistemic uncertainty, the
reducible uncertainty in the learned closure due to limited training data, from
aleatoric uncertainty, the irreducible scatter of the high-fidelity target at
given inputs. Galilean and rotational invariance, symmetry, and tracelessness are
enforced through the tensor-basis representation, while physical realisability
is treated separately through a soft regularisation term.
A controlled regression benchmark is first used to assess predictive accuracy,
calibration, and the choice of Bayesian parameterisation. Placing the posterior
on the spline coefficients gives the most favourable overall balance of
accuracy, calibration, parameter cost, and interpretability. Applied to
Reynolds-stress anisotropy, the BKAN is accurate in distribution and its
uncertainty is both calibrated and informative, with larger uncertainty tending
to identify larger prediction errors. Under geometric extrapolation, however,
the model becomes overconfident: epistemic uncertainty increases but does not
fully account for the growth in prediction error. Finally, the framework is
extended to three-dimensional wind-farm wakes, where the learned correction
transfers to unseen layouts and posterior samples are propagated through the
RANS solver---with the aleatoric component injected as a spatially correlated
field---to obtain uncertainty bounds on the predicted wake flow.

\end{abstract}


\end{frontmatter}


\section{Introduction} \label{sec:Introduction}

Reynolds-averaged Navier--Stokes (RANS) modelling remains a practical basis for
turbulent flow prediction when repeated high-fidelity simulations are
computationally prohibitive. A central limitation, however, lies in the closure
of the Reynolds stresses. Most commonly used models rely on the linear eddy
viscosity assumption, which relates the Reynolds-stress anisotropy directly to
the mean strain rate through the Boussinesq hypothesis. This representation is
computationally efficient and numerically robust, but its constitutive form
cannot reproduce important stress responses that arise in flows with separation,
strong streamline curvature, or secondary motion. The resulting modelling
errors are therefore often structural: adjusting closure coefficients may alter
the prediction, but cannot recover physical behaviour that is absent from the
model form itself \cite{eidi2026surrogate}.

Data-driven turbulence closures have been developed to address some of these
structural limitations. Using high-fidelity data, such approaches can learn
corrections to the Reynolds-stress anisotropy or to the production of turbulent
kinetic energy that are not represented by conventional models. Tensor-basis
neural networks \cite{ling2016reynolds} embed Galilean and rotational invariance
into a learned anisotropy mapping. Field inversion and machine learning
approaches \cite{parish2016paradigm} infer spatially varying model corrections,
while sparse symbolic regression methods
\cite{weatheritt2016novel,schmelzer2020discovery} seek explicit algebraic
closures from data. These approaches have improved both \textit{a priori} and
\textit{a posteriori} predictions for a range of canonical turbulent flows.
Broader perspectives on data-driven turbulence modelling and RANS model
uncertainty are given by \cite{duraisamy2019turbulence} and
\cite{xiao2019quantification}, respectively.

Despite this progress, two limitations remain important for the practical use
of learned closures. The first is interpretability. Neural closures can be
difficult to inspect, making it hard to determine which relationships have been
learned, whether these relationships are physically meaningful, and how they
behave outside the training regime. The second is predictive confidence.
Deterministic closures provide point predictions without indicating how reliable
those predictions are when the model is applied to conditions that differ from
those represented in the training data.

\begin{figure}[htbp]
  \centering
  \includegraphics[width=\textwidth]{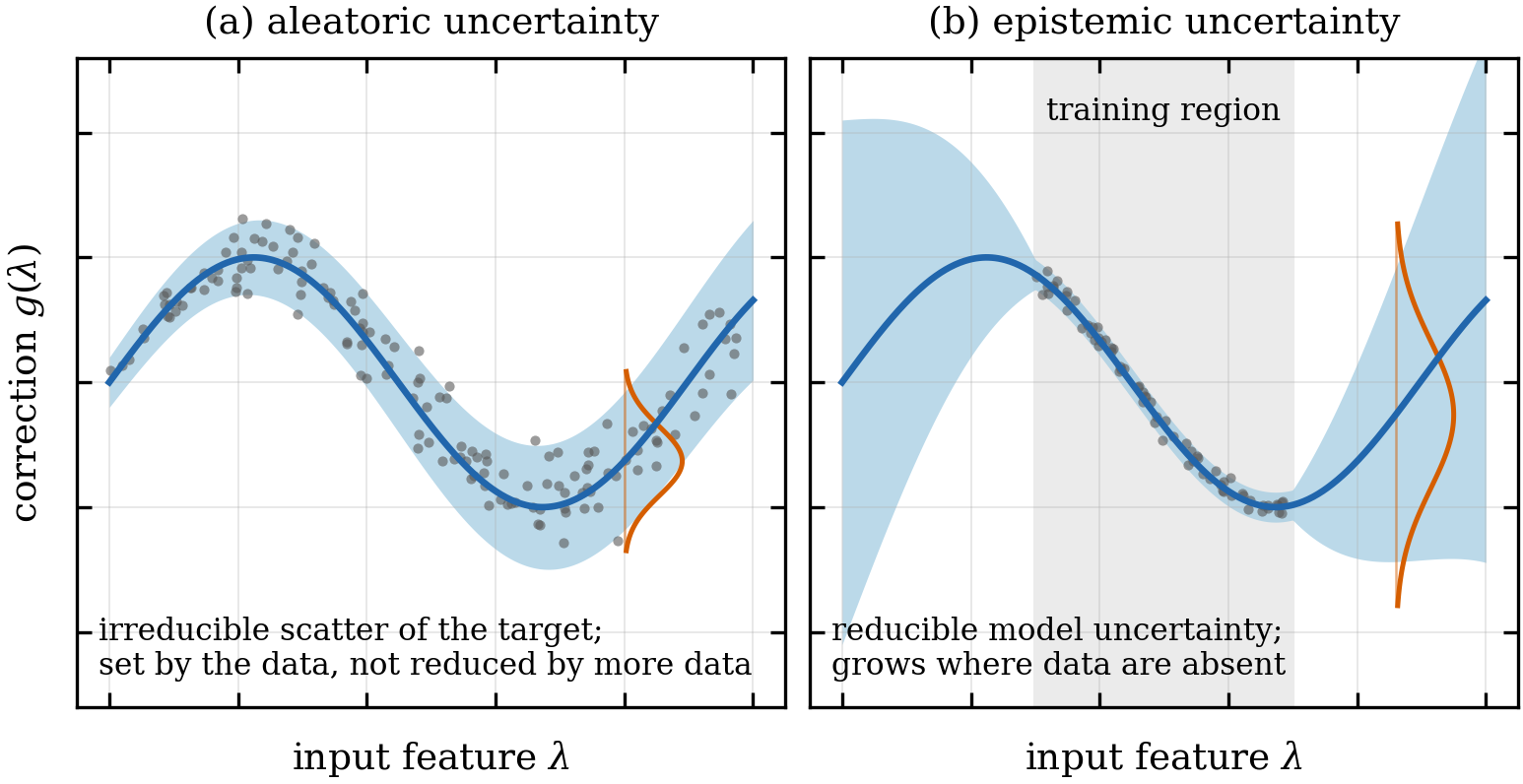}
  \caption{The two kinds of predictive uncertainty for a data-driven closure,
  sketched for a scalar correction $g(\lambda)$ against an input feature.
  (a)~Aleatoric: the target scatters about the mean at fixed input; the
  predictive band follows this scatter and is not reduced by more data.
  (b)~Epistemic: data constrain the model only over part of the input range,
  so the band is tight there and widens where data are absent.}
  \label{fig:uncertainty_concept}
\end{figure}

When a learned closure does report uncertainty, it is useful to distinguish its
two sources, because they carry different information for a turbulence model and
behave differently at deployment, as sketched in
Figure~\ref{fig:uncertainty_concept}.
\emph{Aleatoric} uncertainty is the irreducible scatter of the high-fidelity
target at a given set of input features: at fixed mean-flow invariants the
reference anisotropy still varies, owing to noise in the data and to physical
dependencies that the features do not capture. It is a property of the data and
does not diminish as more data of the same kind are gathered. \emph{Epistemic}
uncertainty, by contrast, is uncertainty in the learned model itself, arising
from finite and unevenly distributed training data; it is small where the data
constrain the closure and grows where they do not, for example on an unseen
geometry, and can in principle be reduced by broadening the training set. For a
RANS closure the distinction is practical: epistemic uncertainty flags when a
prediction is an extrapolation and should be trusted less, whereas aleatoric
uncertainty sets a floor on how closely any closure of this form can match the
reference. In this work we make the separation explicit and quantify both
components, defining them formally in Section~\ref{sec:method}.

These two issues have largely been studied separately. Bayesian and ensemble
closures \cite{geneva2019quantifying}, together with physics-based eigenspace
perturbation methods \cite{emory2013modeling,eidi2022data}, provide estimates of
model uncertainty, but learned closures with uncertainty quantification often
remain difficult to interpret. Moreover, their uncertainty is commonly assessed
primarily through interval coverage. Coverage alone does not show whether larger
predicted uncertainty corresponds to larger model error, or whether the
uncertainty remains reliable under extrapolation. Conversely, symbolic methods
and, more recently, Kolmogorov--Arnold turbulence closures
\cite{mcconkey2025realisability,kalia2025kolmogorov} provide more inspectable
model forms, but are generally deterministic and therefore do not quantify
predictive confidence. In particular, \cite{kalia2025kolmogorov} identifies
uncertainty quantification within the tensor-basis Kolmogorov--Arnold framework
as an open problem.

An important precedent is the sparse Bayesian learning of explicit algebraic
Reynolds-stress models \cite{cherroud2022sparse}. In that approach,
frame-invariant tensor polynomials provide an interpretable closure form, while
a posterior distribution over the model coefficients provides uncertainty
quantification that can also be propagated through the flow solver. The
functional form, however, is selected from a prescribed dictionary of candidate
tensor terms, and the Bayesian treatment is applied to the coefficients of a
model that is linear in the selected features. A remaining question is whether
the constitutive functions themselves can be learned while retaining an
inspectable model structure and a predictive distribution. The resulting
uncertainty should also be assessed using criteria that go beyond nominal
interval coverage.

Kolmogorov--Arnold networks (KANs) \cite{liu2025kan} offer a useful model
structure for this purpose. Unlike standard neural networks, which combine
learned linear transformations with fixed nonlinear activation functions, KANs
place learnable univariate functions on the network edges. These functions are
typically represented by splines and can be inspected directly or, when their
structure is sufficiently simple, approximated by symbolic expressions. This
has motivated several interpretable and symbolic KAN formulations for scientific
machine learning \cite{faroughi2026kolmogorov,faroughi2026symbolic}. The same
property is attractive for tensor-basis turbulence closures, where the learning
problem reduces to determining a relatively small set of scalar coefficient
functions of invariant flow features.

Standard KANs, however, remain deterministic. Like conventional multilayer
perceptrons (MLPs), they provide point predictions without quantifying
predictive uncertainty. A Bayesian treatment addresses this limitation by
assigning posterior distributions to selected model parameters. These
posteriors induce distributions over the learned edge functions and, in turn,
over the model predictions, allowing epistemic uncertainty associated with the
learned closure to be distinguished from aleatoric uncertainty associated with
the data. Bayesian variants of KANs have recently been introduced in the machine
learning literature \cite{hassan2025bayesian,gao2025scalable}, but their use as
physically structured turbulence closures has not yet been examined. It also
remains unclear which KAN parameters should carry the posterior and how the
resulting uncertainty behaves when the model is applied outside the distribution
represented by the training data.

The Bayesian Kolmogorov--Arnold network (BKAN) developed in this work extends
the deterministic KAN in the same sense that a Bayesian neural network (BNN)
extends a conventional MLP. BNNs have previously been used for turbulence
closures with uncertainty quantification
\cite{geneva2019quantifying,buchanan2026bayesian}, but their internal mappings
remain difficult to interpret. Here, we combine the probabilistic treatment of
Bayesian neural closures with the inspectable edge functions of KANs and the
invariant structure of a tensor-basis representation.

Specifically, we formulate a Bayesian tensor-basis KAN for Reynolds-stress
anisotropy that combines interpretable spline functions with a posterior
predictive distribution, while Galilean and rotational invariance, symmetry,
and tracelessness are enforced through the tensor-basis representation. We
examine whether the posterior should be placed on the spline coefficients, the
edge amplitudes, or both, and compare these choices in terms of calibration,
parameter count, and interpretability. The quality of the predicted uncertainty
is assessed not only through calibration, but also through sharpness and
sparsification-based error ranking. This provides a more direct test of whether
regions assigned high uncertainty are also regions where the model makes larger
errors, and how this relationship changes under extrapolation. The closure is
then evaluated across a hierarchy of turbulent flows, from canonical planar and
duct configurations to three-dimensional wind farm wakes. For the wind farm
cases, posterior samples are also propagated through the flow solver to quantify
uncertainty in predicted flow quantities.

The remainder of the paper is organised as follows.
Section~\ref{sec:method} presents the BKAN closure, its tensor-basis formulation,
and the decomposition of predictive uncertainty, with the full variational
formulation given in \ref{app:vi}. Section~\ref{sec:benchmark} evaluates the
underlying Bayesian regressor on a controlled benchmark.
Section~\ref{sec:anisotropy} considers Reynolds-stress anisotropy prediction in
both in-distribution and cross-geometry settings. Section~\ref{sec:wakes}
extends the formulation to three-dimensional wind farm wakes and
solver-propagated uncertainty. Section~\ref{sec:discussion} discusses the
implications, the trust boundary, and the limitations of the approach.
Section~\ref{sec:conclusions} summarises the main conclusions.

\section{A Bayesian Kolmogorov--Arnold network for tensor-basis closures}
\label{sec:method}

\subsection{The closure problem and tensor-basis representation}
\label{sec:method:tb}

The Reynolds-stress anisotropy tensor is defined as
\begin{equation}
  b_{ij} = \frac{\overline{u_i u_j}}{2k} - \frac{1}{3}\delta_{ij},
  \qquad
  k = \tfrac12\,\overline{u_i u_i},
  \label{eq:bij}
\end{equation}
Here $\overline{u_i u_j}$ is the Reynolds-stress tensor, $u_i$ the fluctuating
velocity, $k$ the turbulent kinetic energy, and $\delta_{ij}$ the Kronecker
delta; $b_{ij}$ is the normalised, trace-free anisotropy tensor obtained by
scaling the Reynolds stress by $2k$ and removing its isotropic part. It is the
closure quantity considered in this work. Linear eddy viscosity
models often represent this anisotropy inadequately because they restrict its
relation to the mean strain rate. Following \cite{pope1975more}, and assuming
that the anisotropy is a local function of the normalised mean strain- and
rotation-rate tensors, $\Shat$ and $\What$, alone, a general frame-invariant
relation between $b_{ij}$ and these tensors can be expressed as a finite tensor
polynomial expansion,
\begin{equation}
  b_{ij}(\bm{x})
  =
  \sum_{n=1}^{N}
  g_n\big(\lambda_1,\dots,\lambda_m\big)\,
  T^{(n)}_{ij}\big(\Shat,\What\big),
  \label{eq:tb}
\end{equation}
where $T^{(n)}_{ij}$ are integrity-basis tensors and $\lambda_p$ are scalar
invariants formed from $\Shat$ and $\What$. For two-dimensional mean flows,
three basis tensors and two invariants are sufficient for a perfectly general
representation. In three dimensions, the full representation contains ten
basis tensors and five invariants. The tensors and invariants used in this work
are given in \ref{app:basis}.

Equation~\eqref{eq:tb} separates the tensorial structure of the closure from
the unknown scalar coefficient functions $g_n$. Since the basis tensors are
constructed from $\Shat$ and $\What$ and the coefficients depend on invariant
scalar quantities, the resulting representation preserves the required frame
invariance while enforcing symmetry and tracelessness by construction. The
learning problem is therefore reduced to determining a relatively small set of
scalar coefficient functions rather than learning the individual components of
$b_{ij}$ independently.

For the learned closure, we supplement the tensor invariants with three
additional Galilean-invariant scalar markers: a measure of the balance between
strain and rotation, a turbulence-intensity measure, and a streamline
pressure-gradient marker. These quantities provide information that is not
contained in the pointwise tensor invariants alone. In what follows,
$\bm{\lambda}\in\mathbb{R}^{m}$ denotes the complete scalar feature vector of
dimension $m$ supplied to the network, consisting of the tensor invariants
together with these additional markers. Their definitions are given in \ref{app:basis}.

\subsection{The Kolmogorov--Arnold architecture}
\label{sec:method:kan}

A KAN represents a multivariate function through compositions of learnable
univariate functions placed on the network edges \cite{liu2025kan}. This
differs from an MLP, where learned linear transformations are combined with
fixed nonlinear activation functions. In a KAN, each edge carries a function
of the form
\begin{equation}
  \phi(x) = w_b\,b(x) + w_s\,\phi_s(x),
  \qquad
  \phi_s(x) = \sum_c \theta_c\,B_c(x),
  \label{eq:edge}
\end{equation}
where $b(x)$ is a fixed basis function, $B_c(x)$ denotes the $c$th B-spline
basis function on a prescribed knot grid, and $\theta_c$ is its trainable
coefficient. The amplitudes $w_b$ and $w_s$ scale the fixed and spline
contributions, respectively.
Because each edge function is one-dimensional, it can be inspected directly
over its input range. When a learned function has a sufficiently simple shape,
it can also be approximated by a closed-form expression. We use this property
for the symbolic distillation in Section~\ref{sec:benchmark}, following the
symbolic regression procedure of \cite{liu2024kan2}.

\begin{figure}[htbp]
  \centering
  \includegraphics[width=\textwidth]{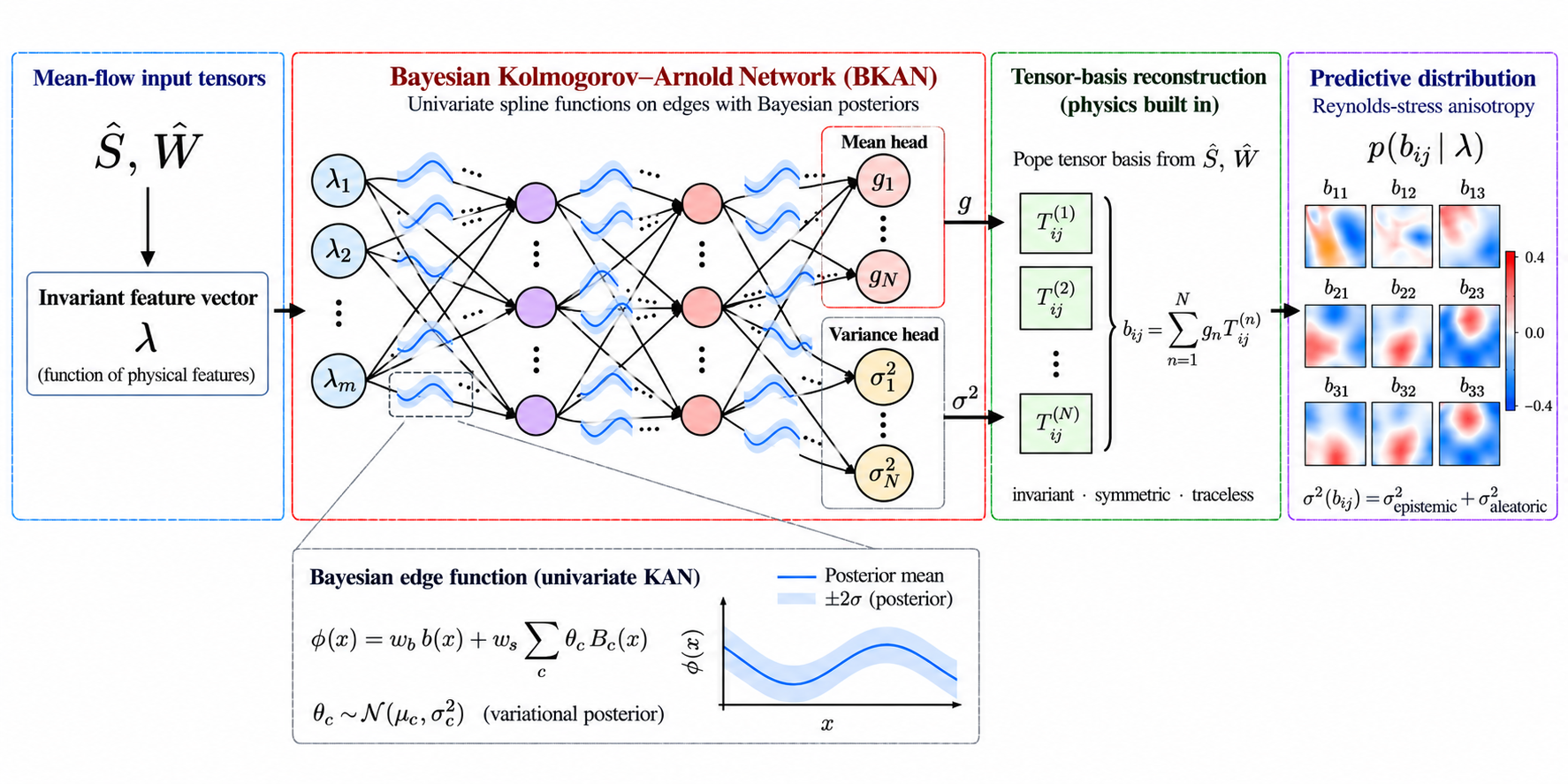}
  \caption{Schematic of the Bayesian tensor-basis KAN closure. The invariant
  feature vector $\bm{\lambda}$ is mapped by a shared KAN with learnable spline
  edges to two output heads: the tensor-basis coefficients $g_n$ and the
  heteroscedastic variance. The coefficients are combined with the fixed Pope
  basis tensors to reconstruct $b_{ij}=\sum_{n=1}^{N}g_nT_{ij}^{(n)}$; posterior
  sampling provides the epistemic uncertainty and the variance head the
  aleatoric uncertainty.}
  \label{fig:schematic}
\end{figure}

In the present closure, the feature vector $\bm{\lambda}$ is mapped by the BKAN
to the scalar tensor-basis coefficients $g_n$. These coefficients are combined
with the fixed Pope basis tensors through \eqref{eq:tb} to reconstruct the
Reynolds-stress anisotropy. A second output head uses the same input features to
predict the input-dependent variance required by the heteroscedastic
likelihood. The resulting architecture, with a shared BKAN representation and
separate coefficient and variance heads, is illustrated in
Figure~\ref{fig:schematic}.

\subsection{Variational Bayesian treatment}
\label{sec:method:vi}

To quantify the epistemic uncertainty in the learned closure---the uncertainty
in the model parameters that arises from training on a finite data set---we place
a mean-field Gaussian variational posterior $q_\psi$ over the subset of model
parameters treated as Bayesian variables, denoted $\bm{\theta}\in\mathbb{R}^{P}$
with $P$ the number of such parameters. The posterior is parameterised by the
variational parameters $\psi=\{(\mu_i,\sigma_i)\}_{i=1}^{P}$, the mean and
standard deviation of each Gaussian factor, which are optimised during training. The model is trained by maximising the evidence lower bound (ELBO),
\begin{equation}
  \mathcal{L}(\psi)
  =
  \underbrace{
    \E_{q_\psi}\!\left[
      \log p(\mathcal{D}\mid\bm{\theta})
    \right]
  }_{\text{expected log-likelihood}}
  -
  \underbrace{
    \KL\!\left(
      q_\psi(\bm{\theta})\,\|\,p(\bm{\theta})
    \right)
  }_{\text{KL divergence}},
  \label{eq:elbo}
\end{equation}
following the Bayes-by-backprop formulation of
\cite{blundell2015weight}. Here, $\KL$ is the Kullback--Leibler divergence
between the variational posterior and the prior.
The local reparameterisation trick \cite{kingma2015variational} is used to
reduce gradient variance through the spline activations. For the spline branch
of an edge with Gaussian coefficients
$\theta_k\sim\mathcal{N}(\mu_k,\sigma_k^2)$, and because the spline $\phi_s(x)$ is
linear in those coefficients, its output at a given input
is itself Gaussian, with mean
$\sum_k B_k(x)\mu_k$ and variance
$\sum_k B_k(x)^2\sigma_k^2$. The spline output can therefore be sampled
directly from this scalar distribution rather than by sampling every
coefficient individually. The full derivation is given in \ref{app:vi}.

The data term is based on a heteroscedastic Gaussian likelihood over the $d$
unique components of the anisotropy tensor. Its negative log-likelihood (NLL)
is
\begin{equation}
  \mathrm{NLL}
  =
  \frac{1}{2}
  \sum_{c=1}^{d}
  \left[
    \log \sigma_c^2(\bm{\lambda})
    +
    \frac{
      \big(b_c-\hat b_c\big)^2
    }{
      \sigma_c^2(\bm{\lambda})
    }
  \right],
  \label{eq:nll}
\end{equation}
where $b_c$ and $\hat b_c$ denote the reference and predicted values of the
$c$th unique anisotropy component, respectively, and
$\sigma_c^2(\bm{\lambda})$ is its predicted aleatoric variance. Allowing this
variance to depend on the input gives the likelihood a spatially varying width
rather than assuming a constant noise level.

In practice, we minimise the negative ELBO together with the realisability
regularisation introduced below,
\begin{equation}
  J(\psi)
  =
  \underbrace{
    \mathrm{NLL}\big(b_{ij}\mid\bm{\lambda}\big)
  }_{\text{data fit}}
  +
  \alpha
  \underbrace{
    R\big(\hat b_{ij}\big)
  }_{\text{realisability}}
  +
  \beta\,
  \frac{\KL(q_\psi\|p)}{N},
  \label{eq:loss}
\end{equation}
where $N$ is the number of training points, $\alpha$ controls the weight of the
realisability penalty, and $\beta$ controls the contribution of the KL term.
During training, $\beta$ is gradually increased from zero, allowing the network
to first fit the data before the posterior is progressively regularised toward
the prior.

The variance of the Gaussian likelihood is predicted by the second output head
of the BKAN using the same feature vector $\bm{\lambda}$. The head predicts an
input-dependent heteroscedastic log variance, which is mapped to the positive
variance $\sigma_c^2(\bm{\lambda})$ used in \eqref{eq:nll}. This allows the
aleatoric uncertainty to vary with the local flow state
\cite{kendall2017uncertainties,nix1994estimating}. In the present setting this
aleatoric component represents the irreducible scatter of the high-fidelity
target that remains once the invariant features are fixed, arising from noise in
the reference data and from physical dependencies not captured by the chosen
feature set. Unlike the epistemic term, it is not reduced by adding more training
data of the same kind, and the heteroscedastic form lets it grow in regions, such
as strong shear or separation, where the feature-to-anisotropy relationship is
intrinsically more scattered. This variance is pointwise: the aleatoric
uncertainty at a cell depends only on that cell's features, and no correlation
between neighbouring cells is modelled at this stage. Such spatial correlation
becomes relevant only when the aleatoric component is propagated through the flow
solver, where it is introduced explicitly
(Section~\ref{sec:wakes:propagation}).

We consider three prior formulations for the Bayesian parameters: a Gaussian
prior with an analytic KL term, the two-component scale-mixture prior of
\cite{blundell2015weight}, and a Gamma hyperprior that provides automatic
relevance determination \cite{mackay1992practical}. Their effect on the results
is small, and the Gaussian prior is therefore used unless stated otherwise.
The corresponding KL expressions and the local reparameterisation derivation
are given in \ref{app:vi}.

Training begins with a short deterministic stage in which the posterior means
are optimised before variational inference. We found this initialisation useful
for resolving sharp spatial variations in the target fields. Further training
details are given in \ref{app:train}, and the complete training and prediction
procedure is summarised in Algorithm~\ref{alg:train}.

\begin{algorithm}[htbp]
\caption{Training and prediction of the Bayesian tensor-basis KAN closure.}
\label{alg:train}
\begin{algorithmic}[1]

\State \textbf{Inputs:} invariant features $\{\bm{\lambda}^{(p)}\}$,
basis tensors $\{\bm{T}^{(n,p)}\}$,
labels $\{b_{ij}^{(p)}\}$;
pre-training epochs $E_0$;
variational epochs $E_1$;
anneal length $E_a$

\State initialise variational means $\bm{\mu}$ and
log-standard-deviations $\bm{\rho}$

\For{$e=1,\dots,E_0$}
  \Comment{deterministic pre-training of the means}

  \State
  $g_n\gets\mathrm{KAN}(\bm{\lambda};\bm{\mu})$;\quad
  $\hat b_{ij}\gets\sum_n g_nT_{ij}^{(n)}$

  \State
  update $\bm{\mu}$ by gradient descent on
  $\left\langle\|\hat b_{ij}-b_{ij}\|^2\right\rangle$

\EndFor

\For{$e=1,\dots,E_1$}
  \Comment{variational inference}

  \State
  $\beta\gets\min(1,e/E_a)$
  \Comment{KL annealing}

  \For{each mini-batch}

    \State
    sample spline activations by local reparameterisation,
    \eqref{eq:localreparam}

    \State
    $\hat b_{ij}\gets\sum_n g_nT_{ij}^{(n)}$;\quad
    $\log\sigma_{ij}^{2}\gets
    \text{variance-head}(\bm{\lambda})$

    \State
    $J\gets
    \mathrm{NLL}
    +\alpha R(\hat b_{ij})
    +\beta\,\KL/N$
    \Comment{\eqref{eq:loss}}

    \State
    update $(\bm{\mu},\bm{\rho})$ and the variance head
    using Adam on $J$

  \EndFor
\EndFor

\State
\textbf{Predict:}
draw $S$ posterior samples $\{g_n^{(s)}\}$,
evaluate $\bar b_{ij}$ at the posterior-mean coefficients,
and compute the epistemic variance
$\operatorname{Var}_s(b_{ij}^{(s)})$
and aleatoric variance
$\E_s[\sigma_{ij}^2]$
via \eqref{eq:totvar}

\end{algorithmic}
\end{algorithm}

\subsection{Choice of Bayesian parameters}
\label{sec:method:which}

The KAN edge in \eqref{eq:edge} contains two groups of parameters that can
be treated probabilistically. The spline coefficients $\theta_c$ determine the
shape of the learned univariate functions, whereas the amplitudes $w_b$ and
$w_s$ scale the fixed and spline contributions. We therefore consider three
Bayesian parameterisations: a posterior over the spline coefficients only
(b-coef), a posterior over the edge amplitudes only (b-scale), and a posterior
over both sets of parameters (b-full).

The three variants differ in parameter count and in how uncertainty is
represented within each edge function. Each Bayesian parameter is represented
by a variational mean and log-standard-deviation. In b-coef, the spline
coefficients $\{\theta_c\}$ are treated as Bayesian variables while $w_b$ and
$w_s$ remain deterministic. In b-scale, the amplitudes are Bayesian and the
spline coefficients remain deterministic, while b-full places a posterior over
both. Consequently, b-coef introduces two variational parameters for each
spline coefficient, b-scale introduces two for each Bayesian edge amplitude,
and b-full includes both contributions.

This distinction also affects interpretability. Since the spline coefficients
determine the shape of each learned univariate function, placing a posterior
over them produces a distribution over the function itself rather than only
over its overall amplitude. Which parameterisation gives the most useful
balance between predictive performance, uncertainty quality, parameter cost,
and interpretability is not clear \textit{a priori}. We therefore compare the
three variants empirically in Section~\ref{sec:benchmark}.

\subsection{Predictive distribution and its decomposition}
\label{sec:method:uq}

At prediction time, the posterior over the coefficient functions induces a
predictive distribution over $b_{ij}$. Drawing $S$ posterior samples
$\bm{\theta}^{(s)}\sim q_\psi$ gives a corresponding set of reconstructed
anisotropy tensors $b_{ij}^{(s)}$ through \eqref{eq:tb}. Their spread
represents epistemic uncertainty, while the input-dependent variance predicted
by the variance head provides the aleatoric component. By the law of total
variance, the predictive variance is decomposed as
\begin{equation}
  \underbrace{\mathrm{Var}\big[b_{ij}\big]}_{\text{predictive}}
  =
  \underbrace{\mathrm{Var}_{s}\big[b_{ij}^{(s)}\big]}_{\text{epistemic}}
  +
  \underbrace{\E_{s}\big[\sigma_{ij}^2(\bm{\lambda})\big]}_{\text{aleatoric}}.
  \label{eq:totvar}
\end{equation}

The epistemic term reflects uncertainty in the learned closure and is expected
to increase when predictions rely on regions of the input space that are poorly
constrained by the training data. The aleatoric term represents the
input-dependent scatter captured by the heteroscedastic likelihood. This
decomposition therefore separates uncertainty associated with the learned model
from that represented by the likelihood.

For the reported point field, we evaluate the model at the posterior-mean
coefficients rather than averaging the $S$ sampled predictions. This avoids
Monte-Carlo noise in the displayed prediction, while the epistemic variance is
estimated from the posterior samples. The corresponding moments are accumulated
online over $S=50$--$100$ draws, avoiding storage of an array of size
$\mathcal{O}(S\!\times\!N)$. The quality of the resulting uncertainty estimate,
including its calibration, sharpness, and ability to identify regions of larger
prediction error, is examined in Section~\ref{sec:anisotropy}.

\subsection{Physical realisability}
\label{sec:method:realiz}

The anisotropy tensor is physically realisable when its eigenvalues remain
within the interval $[-\tfrac13,\tfrac23]$, or equivalently when the
corresponding barycentric coordinates are non-negative
\cite{banerjee2007presentation}. To discourage predictions outside this
admissible region, we introduce the differentiable penalty
\begin{equation}
  R(b_{ij})
  =
  \big\langle
    \max(0,\lambda_{\max}-\tfrac23)^2
    +
    \max(0,-\tfrac13-\lambda_{\min})^2
  \big\rangle,
  \label{eq:realiz}
\end{equation}
where the average is taken over the training points and
$\lambda_{\min}$ and $\lambda_{\max}$ denote the smallest and largest
eigenvalues of the predicted anisotropy tensor. The penalty follows the
realisability treatment of \cite{mcconkey2025realisability}.

Geometrically, the ordered eigenvalues determine a point in the barycentric
representation, with realisable anisotropy states lying inside the admissible
triangle. Equation~\eqref{eq:realiz} penalises predictions whose extreme
eigenvalues leave the admissible interval, with the strength of the
regularisation controlled by $\alpha$. Importantly, the tensor-basis
representation enforces invariance, symmetry, and tracelessness, but does not
by itself guarantee physical realisability. The penalty is therefore used as a
soft regulariser rather than a hard constraint. Its empirical effect is examined
in Section~\ref{sec:anisotropy} and discussed further in
Section~\ref{sec:discussion}. The penalty is applied only to the bounded
anisotropy target and not to the signed correction considered in
Section~\ref{sec:wakes}.

\subsection{Reference models}
\label{sec:method:baselines}

To isolate the effect of the Kolmogorov--Arnold architecture, we compare the BKAN
with a matched tensor-basis BNN baseline. The BNN uses a mean-field variational
MLP to map the same feature vector $\bm{\lambda}$ to the tensor-basis
coefficients $g_n$, followed by the same fixed tensor-basis reconstruction in
\eqref{eq:tb}. It also uses the same heteroscedastic variance head,
realisability penalty, and training protocol. The two models therefore differ
primarily in the function approximator used to predict the coefficients $g_n$:
a KAN in the proposed model and an MLP in the BNN baseline.
The BNN baseline follows the Bayesian tensor-basis closure framework of
\cite{geneva2019quantifying}. Related Bayesian approaches to turbulence-model
correction have been developed by \cite{cherroud2022sparse} and, concurrently
with the present work, by \cite{buchanan2026bayesian}. For the controlled
regression study in Section~\ref{sec:benchmark}, we additionally compare the
BKAN with a deterministic MLP, a deterministic KAN, Monte Carlo dropout
\cite{gal2016dropout}, and deep ensembles
\cite{lakshminarayanan2017simple}, using matched parameter budgets.

\section{Evidence from controlled regression}
\label{sec:benchmark}

Before applying the BKAN to turbulence modelling, we first assess its behaviour
as a Bayesian regressor on a standard benchmark. The purpose is to determine
whether the underlying model can achieve competitive predictive accuracy at a
comparable parameter budget, provide well-calibrated uncertainty estimates, and
retain the interpretability associated with the KAN structure. We use nine
datasets from the regression benchmark of
\cite{hernandez2015probabilistic}, which is widely used for evaluating
probabilistic neural-network models.

The complete experimental protocol, dataset descriptions, and per-dataset
results are given in \ref{app:uci}. The comparison is performed at matched
parameter budget: for each case, the KAN width is selected such that its number
of trainable parameters does not exceed that of the reference BNN. This reduces
the extent to which performance differences can be attributed simply to model
size. We report three metrics. The root-mean-square error measures predictive
accuracy,
\begin{equation}
  \mathrm{RMSE}=\sqrt{\frac{1}{N}\sum_{i=1}^{N}\big(y_i-\hat y_i\big)^2},
  \label{eq:rmse}
\end{equation}
with $\hat y_i$ the predictive mean. The test log-likelihood measures the quality
of the full predictive distribution,
\begin{equation}
  \mathrm{LL}=\frac{1}{N}\sum_{i=1}^{N}
  \log\mathcal{N}\!\big(y_i\mid\hat y_i,\sigma_i^2\big),
  \label{eq:ll}
\end{equation}
rewarding predictions that are accurate and appropriately uncertain, with
$\sigma_i^2$ the predictive variance. Calibration is measured by the expected
calibration error: for nominal central-interval levels $\{p_k\}_{k=1}^{K}$ with
empirical coverages $\hat p_k$ (the fraction of test targets inside the
corresponding predictive interval),
\begin{equation}
  \mathrm{ECE}=\frac{1}{K}\sum_{k=1}^{K}\big|\hat p_k-p_k\big|,
  \label{eq:ece}
\end{equation}
following \cite{kuleshov2018accurate}, so that $\mathrm{ECE}=0$ corresponds to
intervals with their nominal coverage. The BKAN is compared with a variational
BNN, a deterministic MLP, a deterministic KAN, Monte Carlo dropout, and deep
ensembles.

\subsection{Accuracy and calibration}

\begin{figure}[htbp]
  \centering
  \includegraphics[width=0.66\textwidth]{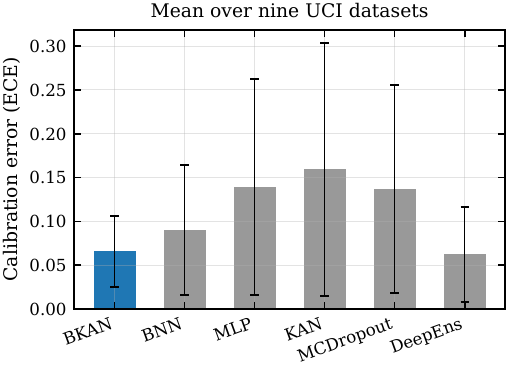}
  \caption{Calibration error (ECE) on the regression benchmark at matched
  parameter budget, averaged over the nine datasets. Predictive error and test
  log-likelihood are near-identical across models and are omitted here (full
  per-dataset results in \ref{app:uci}). The BKAN is among the best calibrated,
  comparable to deep ensembles.}
  \label{fig:uci}
\end{figure}

At matched parameter budget, the BKAN achieves predictive accuracy and test
log-likelihood comparable to those of the BNN across the benchmark suite, while
also providing well-calibrated uncertainty estimates, shown in
Figure~\ref{fig:uci}. The comparison does not indicate uniform superiority:
on several datasets the two Bayesian models perform similarly, while on others
the BNN gives lower predictive error. The full per-dataset results are reported
in \ref{app:uci}.

Averaged over the nine datasets, the BKAN is among the two best-calibrated of
the six models considered, with a mean calibration error of approximately
$0.07$. This is comparable to the deep-ensemble result and lower than the BNN
value of approximately $0.09$, while using a matched or smaller parameter
budget. Predictive error and test log-likelihood are also comparable to the BNN
on most datasets. For the purpose of the present study, these results show that
the BKAN provides competitive regression accuracy together with calibrated
predictive uncertainty. Its behaviour on the turbulence cases can therefore be
examined without the benchmark performance itself being a limiting factor.

\subsection{Bayesian parameterisation study}

\begin{figure}[htbp]
  \centering
  \includegraphics[width=\textwidth]{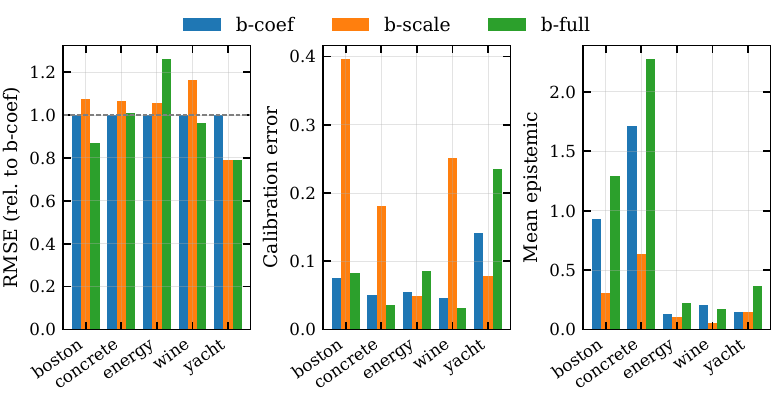}
  \caption{Effect of the Bayesian parameterisation across five benchmark
  datasets. The edge-amplitude posterior (b-scale) gives the least reliable
  calibration, while the spline-coefficient posterior (b-coef) provides the
  most favourable balance between calibration, predictive accuracy, parameter
  cost, and interpretability. The full posterior (b-full) generally produces
  the largest epistemic spread, but at a higher parameter cost.}
  \label{fig:whichparams}
\end{figure}

The benchmark also addresses the design question introduced in
Section~\ref{sec:method:which}: which KAN parameters should carry the posterior.
Figure~\ref{fig:whichparams} compares the three parameterisations across five
benchmark datasets. Placing the posterior only on the edge amplitudes
(b-scale) gives the least reliable calibration, with calibration errors reaching
approximately $0.4$ on some datasets, without providing a consistent improvement
in predictive accuracy. This suggests that uncertainty in the per-edge amplitudes
alone is not sufficient to represent the predictive uncertainty of the learned
functions.

Placing the posterior on the spline coefficients (b-coef) gives the most
favourable overall balance. It remains well calibrated across the datasets and
places the posterior directly on the parameters that determine the shape of each
learned edge function. The full parameterisation (b-full), in which both the
spline coefficients and edge amplitudes are Bayesian, generally produces the
largest epistemic spread. This additional flexibility comes at a higher
parameter cost, without a consistent improvement in calibration over b-coef.

Based on these results, we use b-coef as the default Bayesian parameterisation
throughout the remainder of the paper. The turbulence-anisotropy ablation in
\ref{app:extra} (Table~\ref{tab:appF:which}) supports the same choice: b-coef
gives the lowest RMSE and calibration error, whereas both alternative
parameterisations produce larger epistemic uncertainty and less favourable
overall predictive performance.

\subsection{Interpretability with uncertainty}

We next examine the interpretability of the Bayesian KAN representation using
the tensor-basis closure for the periodic-hill flow. Each edge function in
\eqref{eq:edge} is a one-dimensional curve that can be inspected directly.
Under the b-coef parameterisation, the posterior over the spline coefficients
also induces a distribution over the learned edge function. Its mean can
therefore be examined together with a posterior uncertainty band and, when the
learned shape is sufficiently simple, approximated by a closed-form expression.

Unlike symbolic KANs that incorporate symbolic structure directly into training
through gated primitive selection \cite{faroughi2026symbolic}, the symbolic
distillation used here is performed \textit{post hoc} on the learned functions.
The recovered expression can therefore be compared directly with the posterior
mean and its uncertainty band.

\begin{figure}[htbp]
  \centering
  \includegraphics[width=0.95\textwidth]{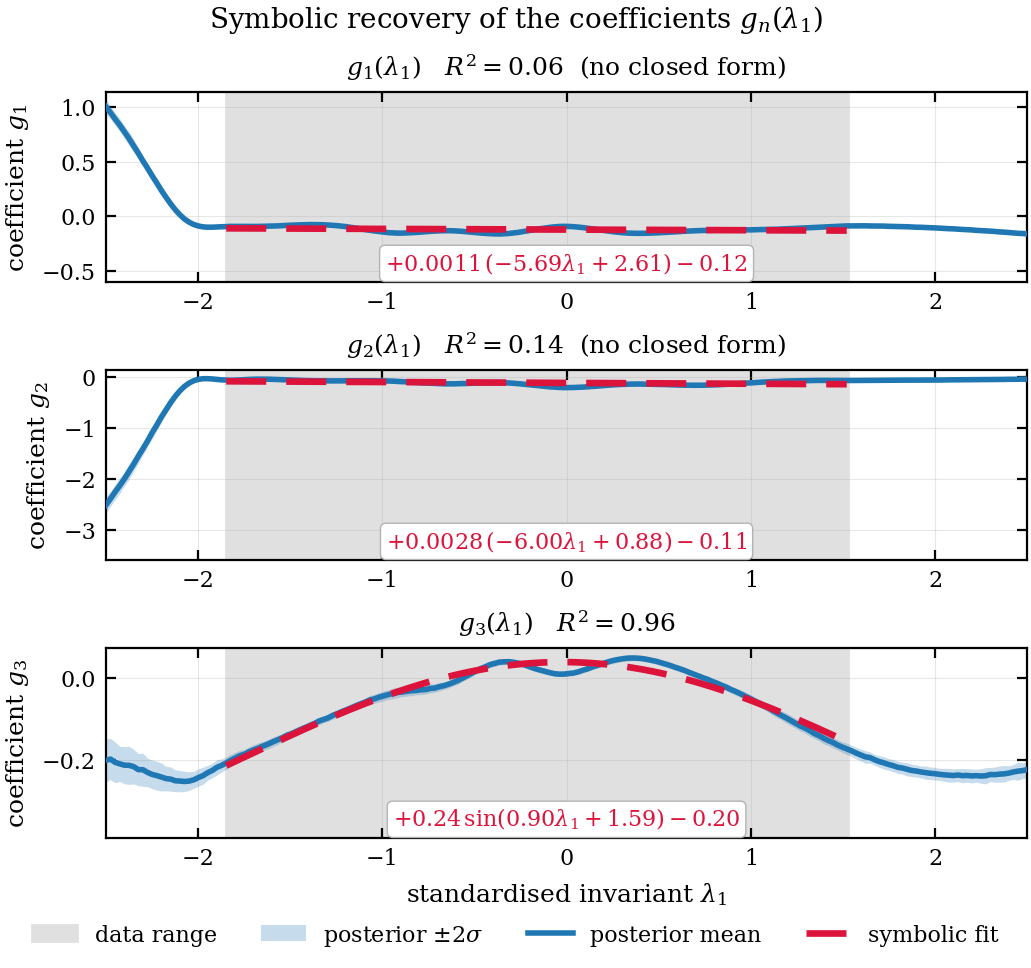}
  \caption{Symbolic recovery of the tensor-basis coefficient functions
  $g_n(\lambda_1)$ for the minimal two-invariant periodic-hill closure, sweeping
  the dominant invariant $\lambda_1$ with $\lambda_2$ held at its median. Each
  panel shows the posterior mean (solid blue), the $\pm2\sigma$ posterior band
  (shaded), and the best closed-form fit (crimson dashed) over the $2$--$98\%$
  data range (grey), within which the candidates are fitted and scored. Only the
  third coefficient admits a simple form; the fits and $R^2$ values are discussed
  in Section~\ref{sec:benchmark}.}
  \label{fig:edge}
\end{figure}

For the minimal two-invariant periodic-hill model used for this analysis, each
candidate expression is fitted and scored only over the range in which the
training data support the learned curve, namely the $2$--$98\%$ band of the
standardised invariant shaded in Figure~\ref{fig:edge}. This restriction is
important. The swept range extends beyond the data, and approximately $87\%$ of
the variance of a coefficient curve lies in the extrapolated tails, so a
candidate that merely follows those tails can reach $R^2\approx0.98$ while
explaining almost nothing where the data constrain the function.

Within the supported range, the third coefficient function is well represented
by
\begin{equation}
  g_3(\lambda_1)
  \;\approx\;
  0.24\,\sin\!\big(0.90\,\lambda_1 + 1.59\big) - 0.20,
  \label{eq:g3symbolic}
\end{equation}
which reproduces the posterior-mean function with $R^2=0.96$. Applying the same
functional form to the upper and lower $\pm2\sigma$ curves gives $R^2=0.95$ in
both cases, so the recovered expression remains consistent across the posterior
band rather than only at its mean. The two remaining coefficients do not admit a
comparably simple closed form over the same range: for both, the best candidate
in the library is linear, with $R^2=0.06$ for $g_1$ and $R^2=0.14$ for $g_2$.

Because the tensor basis is partly collinear in two dimensions, the allocation
of the predicted anisotropy among the individual coefficient functions is not
unique, and the distillation therefore depends on the initialisation. Repeating
the analysis over thirteen random seeds gives thirty-nine coefficient fits, of
which eight reach $R^2>0.9$ while twenty-eight fall below $R^2=0.3$. The leading
coefficient never admits a compact form, its largest $R^2$ over all seeds being
$0.33$, whereas the third distils successfully for six of the thirteen seeds.
Among the successful fits the recovered form is itself divided between
sinusoidal and quadratic candidates. Figure~\ref{fig:edge} therefore shows a
representative case (random seed~17) and should be read as an inspection of one
learned coefficient representation rather than as a unique symbolic form of the
complete closure. The result demonstrates that the internal scalar functions of
the KAN can be inspected together with their posterior uncertainty and, where
their structure is sufficiently simple, represented by compact analytical
expressions, while also delimiting how often that simplicity is available.

\section{Reynolds-stress anisotropy with quantified uncertainty}
\label{sec:anisotropy}

We now apply the proposed closure to its intended task: \textit{a priori}
prediction of the Reynolds-stress anisotropy from RANS mean-flow inputs, using
high-fidelity data as reference. We use the curated dataset of
\cite{mcconkey2021curated}, which provides matched RANS and direct numerical
simulation (DNS) or large eddy simulation (LES) fields for a family of
turbulent flows; the cases and their train/test splits are detailed in
\ref{app:turbcases}. The planar cases considered here are periodic hills, a
converging--diverging channel, and a curved backward-facing step, which involve
separation and streamline curvature. The dataset also includes a square duct,
where secondary flow of Prandtl's second kind is driven by Reynolds-stress
anisotropy and cannot be reproduced by a linear eddy-viscosity model.

For the planar flows, the three-term tensor basis is exact, while the square
duct is represented using the full ten-term basis. Throughout this section, we
distinguish between two prediction regimes. By a \emph{flow family} we mean a
group of flows that share the same geometry type and governing physics and
differ only through a parameter such as the Reynolds number or a geometric
ratio. \textit{In distribution} then refers to interpolation within such a
family, where the training and test data are held-out cells, Reynolds numbers,
or parameter values drawn from the same family. \textit{Out of distribution}
refers to prediction on a geometry that belongs to no family represented in
training and therefore provides a more demanding test of generalisation. The architecture,
input features, and training procedure are summarised in \ref{app:train}. All
in-distribution metrics are reported as means over three random seeds.

\subsection{Accuracy in distribution}
\label{sec:aniso:indist}

\begin{figure}[htbp]
  \centering
  \includegraphics[width=\textwidth]{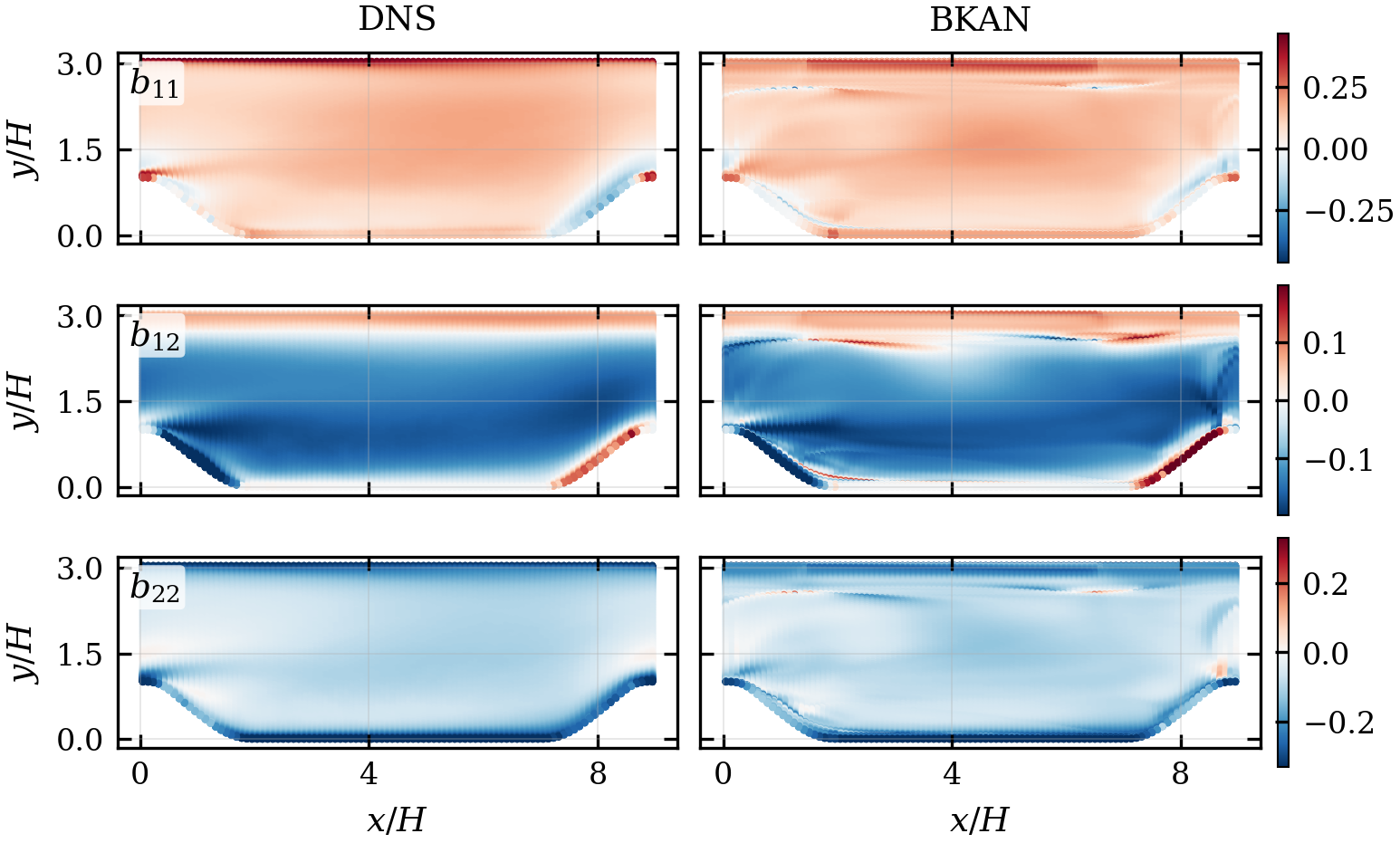}
  \caption{In-distribution anisotropy prediction for the periodic-hill flow:
  DNS reference and BKAN prediction for selected components. The predicted
  fields reproduce the main structures through the separated and recirculating
  regions.}
  \label{fig:phills}
\end{figure}

For the periodic-hill flow, the closure predicts the Reynolds-stress
anisotropy with an RMSE of $0.046\pm0.001$ over three random seeds. The small
variation between seeds indicates limited sensitivity to the model
initialisation. The predicted component fields reproduce the main spatial
structures of the DNS reference through the separated shear layer and
recirculation region, with the largest discrepancies remaining locally around
the more strongly varying parts of the flow, shown in
Figure~\ref{fig:phills}.

\begin{figure}[htbp]
  \centering
  \includegraphics[width=\textwidth]{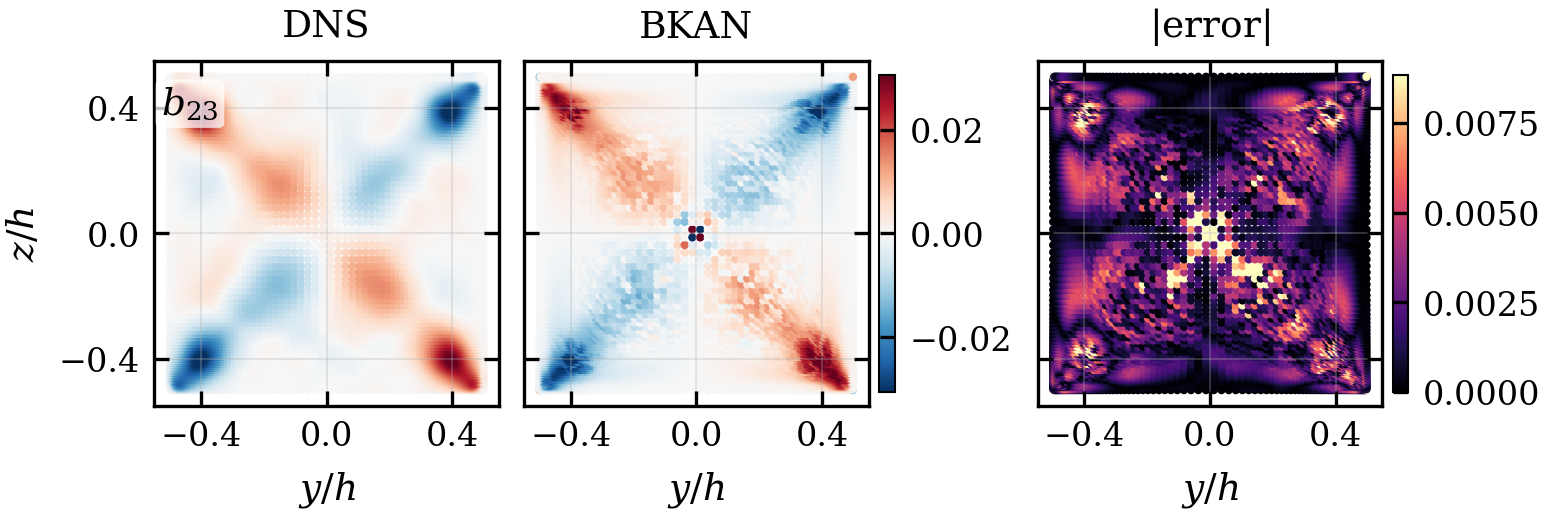}
  \caption{In-distribution prediction for the square duct (ten-term basis),
  showing the shear component $b_{23}$ that drives the secondary motion: DNS
  reference, BKAN prediction, and pointwise absolute error.}
  \label{fig:duct}
\end{figure}

The square duct provides a different test because its secondary flow is driven
by Reynolds-stress anisotropy. In particular, gradients of the cross-plane
normal-stress difference together with the shear component $b_{23}$ contribute
to the generation of streamwise vorticity. These effects cannot be represented
by a linear eddy-viscosity model, which constrains the anisotropy to be
proportional to the mean strain rate. The BKAN reproduces the spatial structure
and sign of the shear component $b_{23}$ that drives the secondary motion, shown
in Figure~\ref{fig:duct}.

\begin{figure}[htbp]
  \centering
  \includegraphics[width=\textwidth]{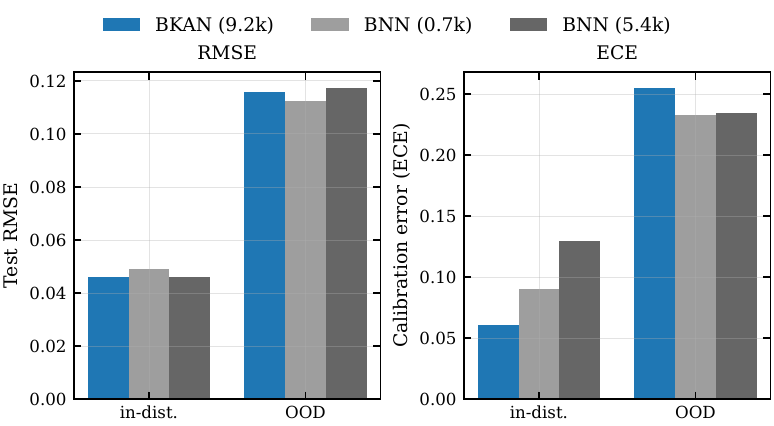}
  \caption{Comparison of the BKAN with two tensor-basis BNN baselines. The
  numbers in parentheses indicate the number of model parameters. All models
  use the same tensor-basis reconstruction, realisability penalty,
  heteroscedastic variance formulation, and training protocol. In distribution,
  the models give comparable predictive accuracy, while the BKAN has the lowest
  calibration error. The out-of-distribution results are discussed in
  Section~\ref{sec:aniso:ood}.}
  \label{fig:bnnbaseline}
\end{figure}

To examine the effect of the Kolmogorov--Arnold architecture, we compare the BKAN
with the tensor-basis BNN introduced in
Section~\ref{sec:method:baselines}. The two models use the same tensor-basis
reconstruction, realisability penalty, heteroscedastic variance head, and
training protocol, and differ primarily in the function approximator used to
map the input features to the coefficients $g_n$. In distribution, the BKAN and
the larger BNN give comparable predictive errors, with RMSE values close to
$0.046$, as shown in Figure~\ref{fig:bnnbaseline}. Their calibration differs
more clearly:
the BKAN has an ECE of approximately $0.06$, compared with approximately $0.09$
for the smaller BNN and $0.13$ for the wider BNN. Increasing the BNN capacity
therefore does not improve its predictive accuracy in this comparison and is
accompanied by poorer calibration. The BKAN achieves comparable accuracy while
retaining the inspectable KAN representation and giving the lowest
in-distribution calibration error among the three models shown.

\subsection{Calibration and informativeness of predictive uncertainty}
\label{sec:aniso:uq}

Predictive accuracy alone does not establish whether the associated uncertainty
estimate is useful. We therefore assess three complementary properties:
calibration, sharpness, and informativeness.

\emph{Calibration} measures whether the predicted intervals have their stated
coverage. A reliability diagram plots the empirical coverage $\hat p_k$ against
the nominal level $p_k$, and the expected calibration error of
Eq.~\eqref{eq:ece} summarises the departure from the diagonal
($\mathrm{ECE}=0$ for intervals with exactly their nominal coverage).

\emph{Sharpness} measures how concentrated the predictive distribution is,
independently of the reference values. We report the mean total predictive
standard deviation,
\begin{equation}
  \mathrm{Sharp}=\frac{1}{N}\sum_{i=1}^{N}\sigma_{\mathrm{tot},i},
  \label{eq:sharp}
\end{equation}
with $\sigma_{\mathrm{tot},i}$ the total predictive standard deviation at point
$i$; smaller values indicate a more concentrated distribution. It is reported
together with calibration because nominal coverage can otherwise be obtained
simply by widening the intervals.

\emph{Informativeness} measures whether the predicted uncertainty ranks the
actual errors. Predictions are sorted by predictive uncertainty and
progressively removed, giving a sparsification curve $e_{\mathrm{unc}}(f)$ of
mean remaining error against the fraction $f$ removed. This is compared with an
oracle curve $e_{\mathrm{oracle}}(f)$ obtained by instead sorting on the true
error; the area between them,
\begin{equation}
  \mathrm{AUSE}=\int_0^1\big[e_{\mathrm{unc}}(f)-e_{\mathrm{oracle}}(f)\big]\,
  \mathrm{d}f,
  \label{eq:ause}
\end{equation}
is the area under the sparsification error. A random ordering provides an upper
reference, and $\mathrm{AUSE}\!\to\!0$ when the uncertainty ranks errors as well
as the true error itself.

\begin{figure}[htbp]
  \centering
  \includegraphics[width=0.92\textwidth]{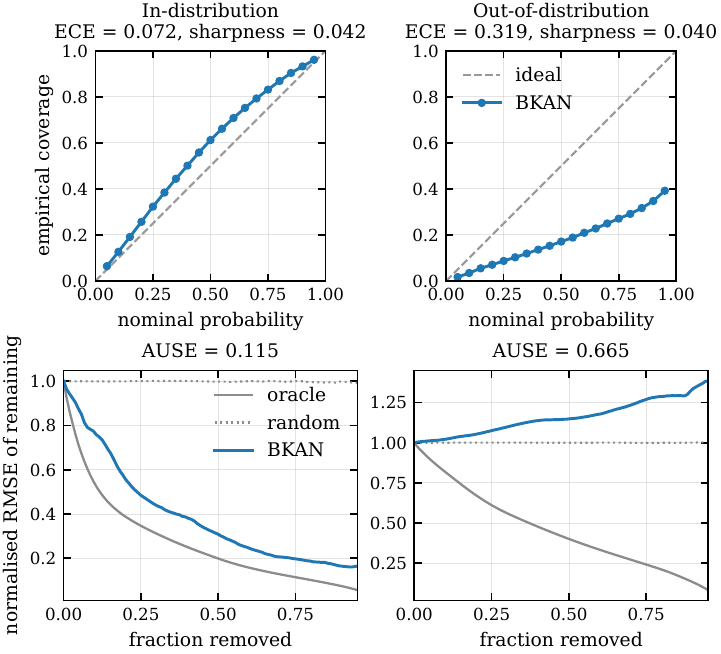}
  \caption{Quality of the predictive uncertainty in distribution (left) and
  out of distribution (right). The top row shows reliability diagrams together
  with the corresponding sharpness, while the bottom row shows sparsification
  curves relative to oracle and random error rankings. In distribution, the
  BKAN is well calibrated and its uncertainty provides an informative ranking
  of prediction error. Out of distribution, calibration and error ranking
  deteriorate substantially; this behaviour is examined further in
  Section~\ref{sec:aniso:ood}.}
  \label{fig:uq}
\end{figure}

In distribution, the reliability curve lies close to the ideal diagonal and is
slightly conservative, with empirical coverage generally exceeding the nominal
coverage (Figure~\ref{fig:uq}, top-left). The corresponding calibration error is
$\mathrm{ECE}=0.072$, with a sharpness of $0.042$. The sparsification result is
also favourable: the BKAN curve remains substantially closer to the oracle than
to the random ordering, with $\mathrm{AUSE}=0.115$
(Figure~\ref{fig:uq}, bottom-left). This indicates that regions assigned larger
uncertainty also tend to contain larger prediction errors.

\begin{figure}[htbp]
  \centering
  \includegraphics[width=\textwidth]{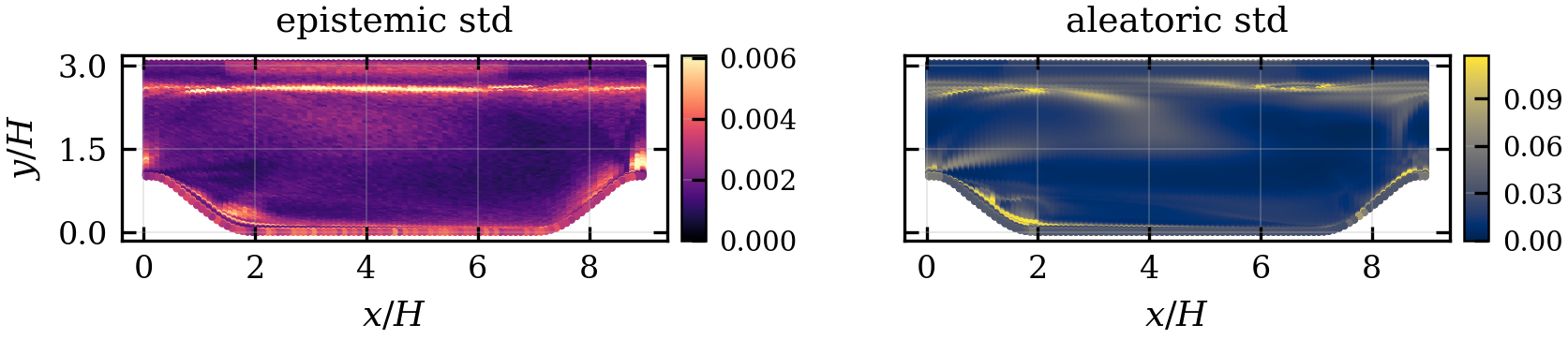}
  \caption{Epistemic (left) and aleatoric (right) predictive standard deviation
  of the $b_{12}$ component for the periodic-hill flow. The epistemic component
  is concentrated along the separated shear layer and near reattachment,
  whereas the aleatoric contribution has a different spatial distribution.}
  \label{fig:phills-unc}
\end{figure}

The spatial structure of the uncertainty provides a complementary view. For the
periodic-hill flow, the epistemic standard deviation of $b_{12}$ is largest
along the separated shear layer and around reattachment in
Figure~\ref{fig:phills-unc}. These are also regions where the learned mapping
is more strongly challenged by the local flow behaviour. The aleatoric and
epistemic contributions have distinct spatial distributions, showing that the
uncertainty estimate contains information beyond aggregate coverage alone.

\subsection{The limits of extrapolation}
\label{sec:aniso:ood}

\begin{figure}[htbp]
  \centering
  \includegraphics[width=\textwidth]{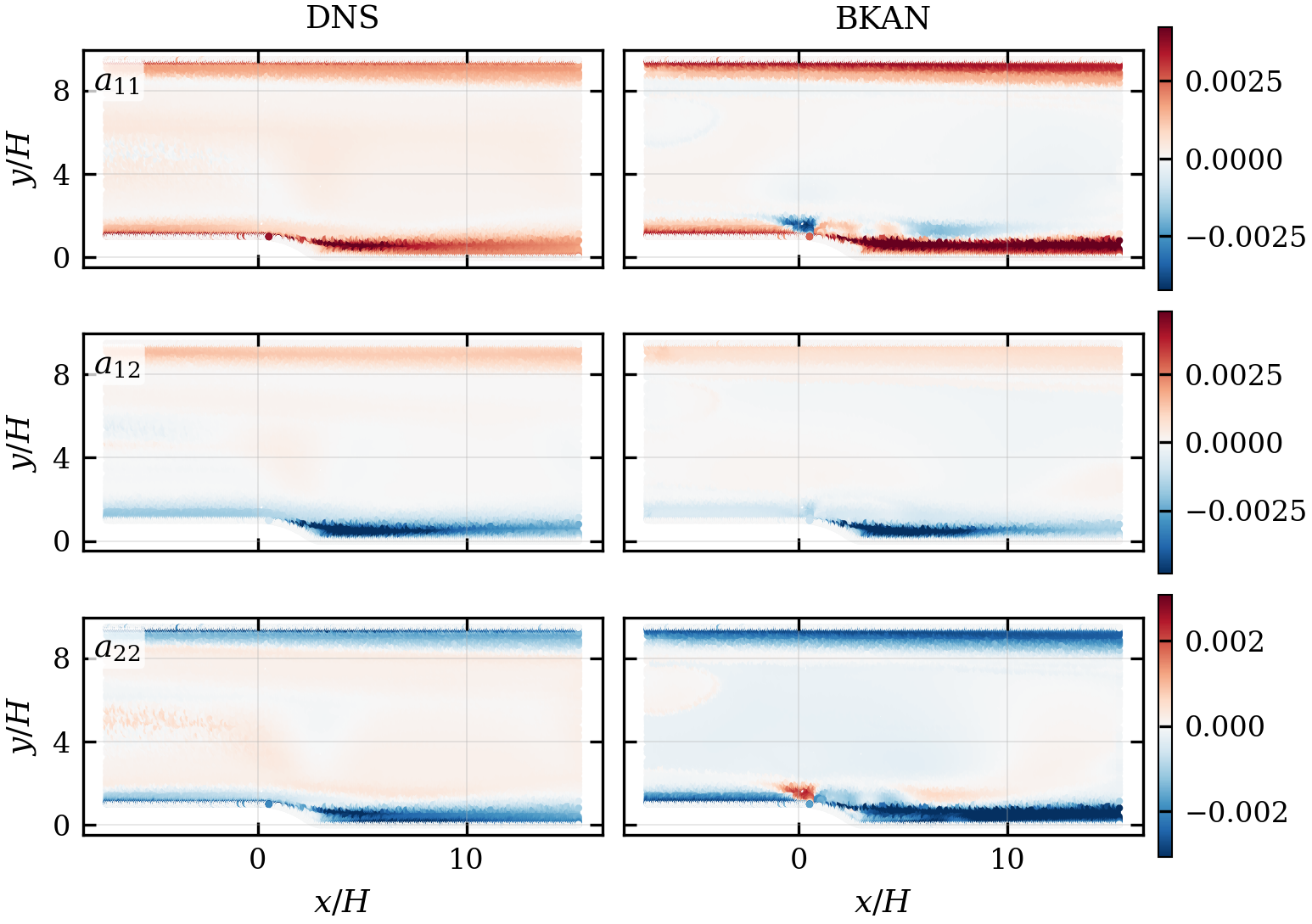}
  \caption{Out-of-distribution prediction for the withheld curved
  backward-facing-step geometry, shown as the unnormalised anisotropy
  $a_{ij}=2k\,b_{ij}$: because the turbulent kinetic energy $k$ is small here,
  the normalised $b_{ij}$ is dominated by noise. DNS reference and BKAN
  prediction for $a_{11}$, $a_{12}$ and $a_{22}$; the shear component
  generalises more successfully than the normal components.}
  \label{fig:ood}
\end{figure}

We next consider prediction on a geometry that is absent from the training
data. When the model is trained on a subset of the available flows and evaluated
on a withheld geometry, the predictive accuracy deteriorates substantially. The
cross-geometry RMSE increases to $0.116\pm0.009$, compared with
$0.046\pm0.001$ in distribution. The larger variation between random seeds
also indicates greater sensitivity to the learned model under extrapolation.
For the curved backward-facing step shown in Figure~\ref{fig:ood}, the shear
anisotropy generalises more successfully than the normal-stress components,
although all components show larger errors than in the in-distribution case.

The uncertainty estimate also deteriorates out of distribution. The reliability
curve falls well below the ideal diagonal, with
$\mathrm{ECE}=0.319$, showing substantial under-coverage of the predictive
intervals (Figure~\ref{fig:uq}, top-right). At the same time, the sharpness
changes only slightly, from $0.042$ in distribution to $0.040$ out of
distribution. The loss of calibration therefore occurs without a corresponding
increase in the width of the predictive distribution. The sparsification result
also deteriorates markedly, with $\mathrm{AUSE}=0.665$, compared with
$0.115$ in distribution (Figure~\ref{fig:uq}, bottom-right). The predicted
uncertainty consequently provides a much poorer ranking of the actual errors
under geometric extrapolation.

\begin{figure}[htbp]
  \centering
  \includegraphics[width=\textwidth]{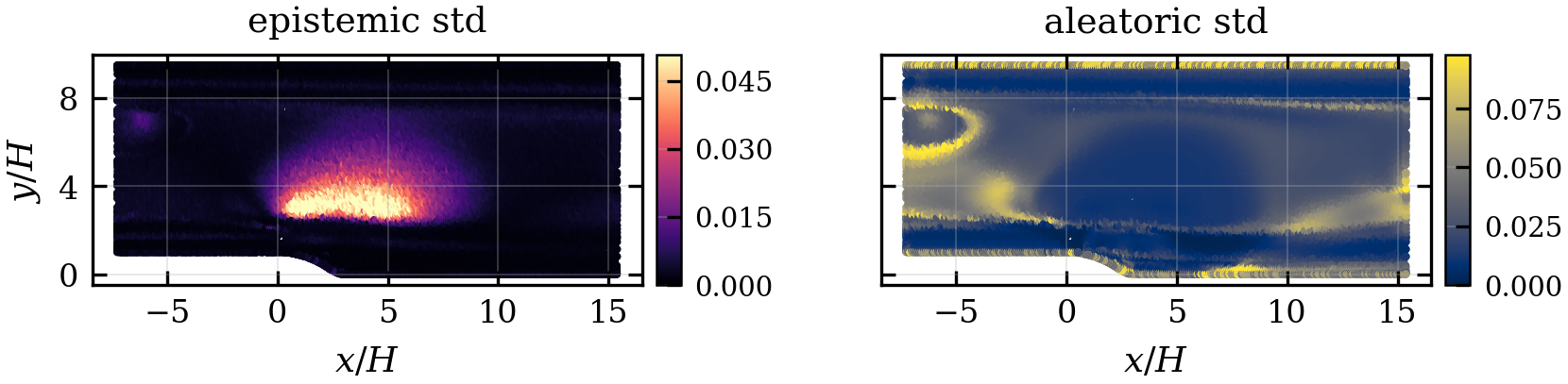}
  \caption{Epistemic (left) and aleatoric (right) predictive standard deviation
  of $b_{12}$ for the withheld geometry. The epistemic component increases and
  concentrates where the geometry departs most from the training flows, while
  the aleatoric component follows the local correction magnitude and is more
  broadly distributed. The epistemic magnitude is nonetheless insufficient to
  account for the prediction error, consistent with the under-coverage in
  Figure~\ref{fig:uq}.}
  \label{fig:ood-unc}
\end{figure}

The spatial epistemic uncertainty nevertheless responds to the change in
geometry. It increases and is concentrated in regions where the withheld flow
differs most strongly from those represented in training, shown in
Figure~\ref{fig:ood-unc}. The aleatoric component, by contrast, follows the
local magnitude of the correction and is more broadly distributed than the
epistemic term, so the two carry complementary information: the epistemic map
flags where the inputs are unfamiliar, whereas the aleatoric map reflects the
intrinsic scatter of the target. The model therefore shows some sensitivity to
extrapolation, but the magnitude of its epistemic uncertainty is insufficient
to account for the resulting prediction error. This combination of increased
epistemic uncertainty and substantial under-coverage defines a practical trust
boundary for the present model: within the represented flow family, the
uncertainty is calibrated and informative, whereas for unseen geometries it
should be interpreted with caution. The implications of this behaviour are
discussed further in Section~\ref{sec:discussion}.

\subsection{Physical realisability of the predictions}
\label{sec:aniso:realiz}

\begin{figure}[htbp]
  \centering
  \begin{subfigure}[t]{0.74\textwidth}
    \centering
    \includegraphics[width=\linewidth]{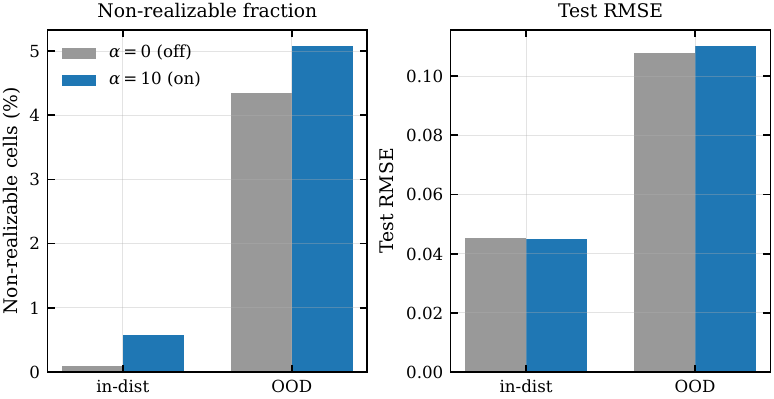}
    \caption{Non-realisable fraction and test error with the penalty off and on.}
  \end{subfigure}\\[6pt]
  \begin{subfigure}[t]{\textwidth}
    \centering
    \includegraphics[width=\linewidth]{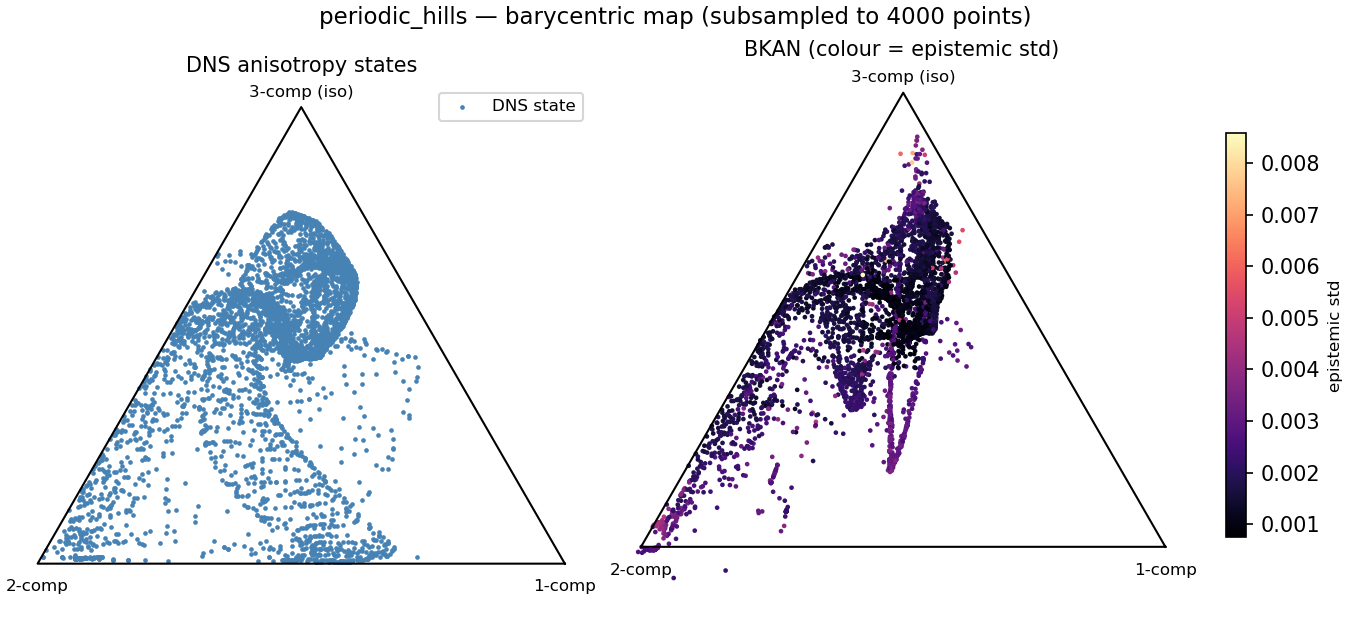}
    \caption{Barycentric map of the predicted anisotropy states.}
  \end{subfigure}
  \caption{Physical realisability of the predicted anisotropy. In distribution,
  most predicted states remain within the realisable barycentric region, while
  the non-realisable fraction increases under geometric extrapolation. At the
  training budget considered here, the soft realisability penalty has no
  material effect on either the non-realisable fraction or the prediction
  error.}
  \label{fig:realiz}
\end{figure}

We finally examine the physical realisability of the predicted anisotropy.
In distribution, the vast majority of predicted states remain within the
realisable barycentric region, with only a small fraction lying outside it,
shown in Figure~\ref{fig:realiz}. The corresponding non-realisable fraction is
$0.09\%$ without the penalty and $0.58\%$ with the penalty. The test RMSE is
essentially unchanged, from $0.0453$ to $0.0451$.

The same conclusion holds more strongly out of distribution. The fraction of
non-realisable predictions increases to a few per cent, reaching $4.35\%$
without the penalty and $5.08\%$ with it, while the RMSE changes from $0.1077$
to $0.1102$. At the training budget considered here, the soft penalty in
\eqref{eq:realiz} therefore does not materially improve realisability or
predictive accuracy. Because the penalty targets the magnitude of the eigenvalue
violations rather than the number of violating cells, and that fraction is
already very small, it can even leave the fraction marginally higher.

These results also clarify the role of the tensor-basis representation. The
basis enforces the required invariance, symmetry, and tracelessness of the
anisotropy tensor, but it does not constrain its eigenvalues to the physically
admissible range. The realisability penalty should therefore be viewed as a
soft regularisation term rather than a guarantee. Enforcing realisability
strictly would require a parameterisation that satisfies the eigenvalue or
barycentric constraints by construction, as discussed further in
Section~\ref{sec:discussion}.

\section{From \textit{a priori} prediction to propagated uncertainty in wind-farm wakes}
\label{sec:wakes}

We next consider a three-dimensional engineering application in which uncertainty
in the turbulence closure can directly affect quantities of practical interest:
wind-farm wakes. Wake deficit and recovery determine the velocity available to
downstream turbines and depend strongly on the turbulent-stress field, which can
be difficult to represent with linear eddy-viscosity models. At the same time,
wind-farm analysis often requires repeated flow evaluations, for which RANS
remains a practical modelling approach. Quantifying the uncertainty associated
with learned RANS corrections is therefore particularly relevant in this setting.
This section first examines the \textit{a priori} prediction of the learned
three-dimensional correction and its transfer between wind-farm configurations,
with the closure evaluated at the reference state at which it is defined, as
Section~\ref{sec:wakes:setup} sets out. We then propagate posterior samples
through the flow solver to determine how the closure uncertainty affects the
predicted wake field.

\subsection{Setup}
\label{sec:wakes:setup}

We consider four atmospheric-boundary-layer cases: an isolated turbine and three
six-turbine arrays with different spacings and aligned or staggered layouts.
Matched high-fidelity reference data are available for all cases
\cite{eidi2021model,eidi2022data}; the configurations and numerical resolutions
are summarised in \ref{app:turbine}. The learned target is the
Reynolds-stress anisotropy correction $b_{ij}^{\Delta}$ relative to the baseline
RANS model. Because these flows are fully three-dimensional, the correction is
represented using the full ten-term tensor basis.

Unlike the anisotropy tensor considered in Section~\ref{sec:anisotropy},
$b_{ij}^{\Delta}$ is a signed correction and is not itself required to satisfy
the realisability bounds imposed on $b_{ij}$. The penalty in
\eqref{eq:realiz} is therefore not applied to the wind-farm correction.
Following \cite{jigjid2025discovery}, the training samples are restricted to the
wake region using a production-based criterion. Cells are retained when the
magnitude of the induced correction to turbulent-kinetic-energy production,
\begin{equation}
  \mathcal{P}^{\Delta}
  =
  -2k\,b_{ij}^{\Delta}\,\partial_j U_i,
  \label{eq:wake_prod_corr}
\end{equation}
exceeds a prescribed threshold. No additional geometric restriction, such as a
wall-distance cut-off, is imposed.
The wake-region criterion is used only to select the training data. At
deployment, the trained closure is evaluated over the entire computational
domain rather than being masked outside the wake. The predicted correction is
comparatively small away from the wake but is retained as predicted instead of
being forced to zero. This avoids introducing an abrupt discontinuity at the
boundary of the training region, while allowing a small non-zero correction in
the surrounding flow. Further details of the wake-region criterion are given in
\ref{app:turbine}.

A further choice concerns the flow state at which the closure inputs are
evaluated. The correction $b^{\Delta}_{ij}$ is defined by a frozen inversion in
which the mean flow is held at the high-fidelity reference, and the invariants
that serve as closure inputs are evaluated on that same solution rather than on
the uncorrected RANS field. This is the consistent choice for a constitutive
relation: the correction is a function of the local state it describes, and
pairing it with invariants taken from the uncorrected solution would instead
learn a map contaminated by the errors of the baseline model, specific to that
model and to the particular flow state it produces. The same convention is not
used in Section~\ref{sec:anisotropy}, where the target is the full anisotropy
tensor and the inputs are the baseline RANS fields, following standard practice
for that task; the distinction follows from the difference between learning a
closure and learning a correction to one.

The consequence for deployment should be stated plainly. Because the inputs are
evaluated on the reference state, the results reported below are conditional on
that state being known, and they therefore measure the quality of the learned
constitutive relation rather than end-to-end predictive performance on a
configuration for which no high-fidelity data exist. On such a configuration the
correction and the mean flow must be brought into agreement iteratively, by
re-evaluating the closure as the solution evolves. The frozen propagation used
in Section~\ref{sec:wakes:propagation} is accordingly a consistency and
uncertainty-propagation study, bounded above by the frozen solution obtained
with the reference correction, which is reported alongside it for exactly this
reason. Coupled evaluation is left to future work.

\subsection{\textit{A priori} prediction and generalisation across layouts}
\label{sec:wakes:apriori}

\begin{figure}[htbp]
  \centering
  \includegraphics[width=\textwidth]{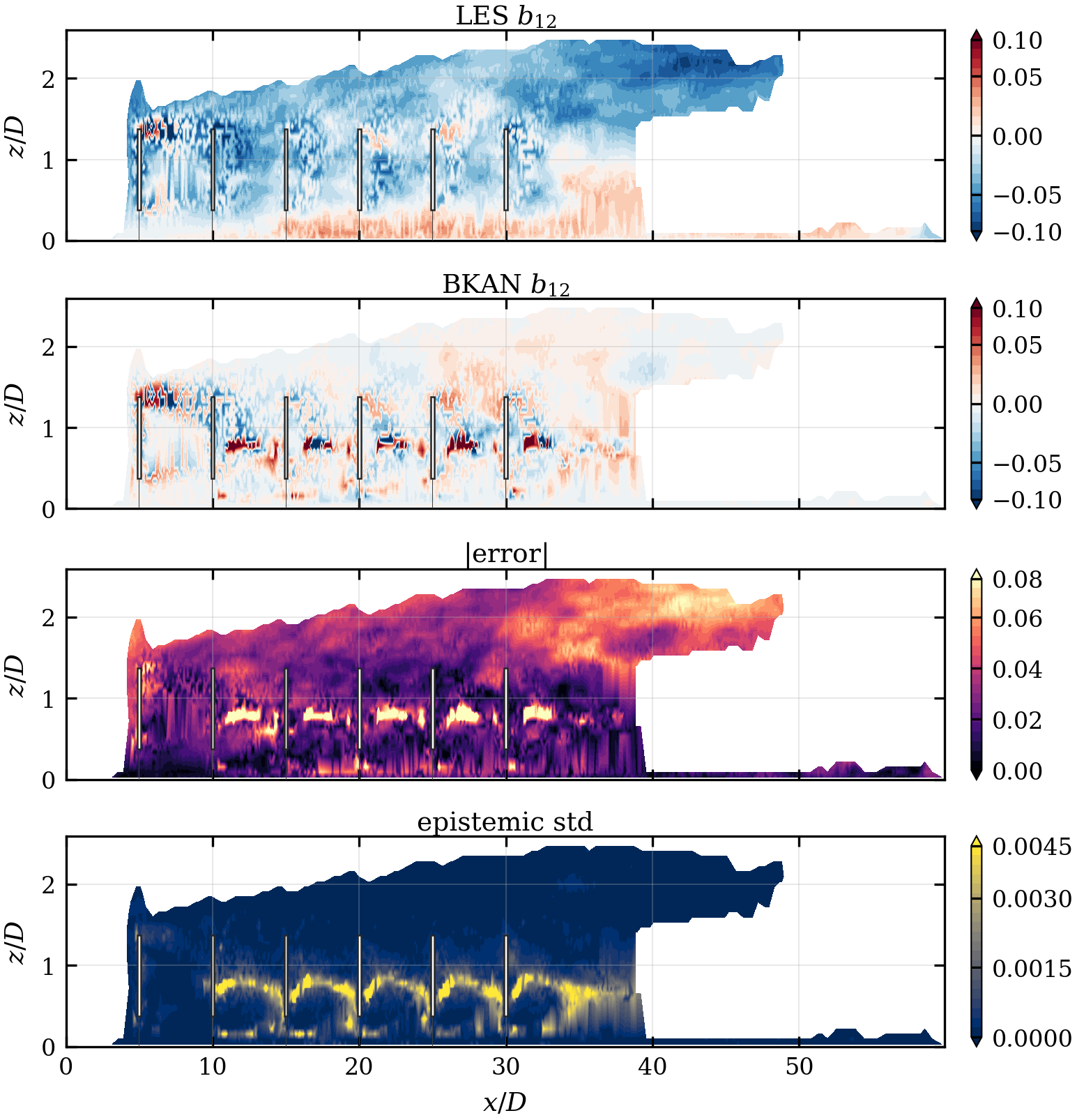}
  \caption{\textit{A priori} prediction of the three-dimensional Reynolds-stress
  anisotropy correction for a six-turbine array, shown on a
  streamwise--vertical plane through the wakes. The high-fidelity reference is
  compared with the BKAN posterior-mean prediction, pointwise absolute error,
  and epistemic standard deviation within the criterion-defined wake region.
  The main spatial structure of the correction is recovered, while the
  epistemic uncertainty is concentrated around the high-shear wake regions.}
  \label{fig:turbine}
\end{figure}

When trained on one six-turbine array and evaluated on the other farm
configurations, the closure transfers well across layouts. For the two withheld
six-turbine cases, the wake-shear anisotropy correction is predicted with
coefficients of determination of $0.87$ and $0.85$ and RMSE values of
$0.03$--$0.04$. These figures quantify how well the correction transfers as a
function of the local state, evaluated at the reference state in both training
and testing; they are not a measure of what a single uncorrected RANS solution
would yield. Figure~\ref{fig:turbine} shows a representative prediction,
including the reference correction, BKAN posterior-mean field, pointwise error,
and epistemic uncertainty. The main spatial structure of the correction is
recovered, while the epistemic uncertainty is concentrated primarily around the
high-shear wake regions.

This transfer between related wind-farm configurations is more successful than
the extrapolation between qualitatively different wall-bounded geometries
considered in Section~\ref{sec:aniso:ood}. The two settings are not directly
comparable in terms of their error metrics, since they involve different target
fields and evaluation domains. Nevertheless, the results suggest that the local
mapping for the wake correction transfers more readily between configurations
that share the same underlying wake physics.

A more demanding test is obtained by training on the isolated single-turbine
case and evaluating the model on the three multi-turbine arrays. In this setting,
the RMSE increases to $0.05$--$0.07$, with coefficients of determination between
approximately $0.4$ and $0.7$, while the calibration error increases to
$0.11$--$0.14$. The isolated-turbine training case does not contain the
wake--wake interactions and merged-wake structures present in the farm
configurations, which provides a plausible explanation for the poorer transfer.
For the subsequent solver propagation, we therefore use a model trained on a
six-turbine array, where the training data already contain the multi-wake
behaviour relevant to the deployment cases.

Because the predicted correction is subsequently propagated through the flow
solver, we also examine the sensitivity of the deployed model to its
architecture. Variations in network width, depth, and spline-grid resolution
produce only small changes in the validation and test errors and in the
predicted correction field. Across these configurations, the field changes by
at most approximately $10\%$ of its magnitude relative to the selected model.
Removing the additional invariant features has the largest effect of the
variations considered, giving both the highest errors and the largest change in
the deployed correction field, although the absolute change remains modest.
The complete sensitivity study is reported in \ref{app:turbine}. Based on this
assessment, we fix the architecture used for the posterior ensemble propagated
through the solver.

\subsection{Propagating the posterior through the solver}
\label{sec:wakes:propagation}

The \textit{a priori} uncertainty in $b^{\Delta}_{ij}$ becomes practically
relevant when it is propagated to quantities predicted by the flow solver.
Moreover, \textit{a priori} accuracy is not necessarily a reliable indicator of
\textit{a posteriori} performance \cite{macart2021embedded}, so the effect of the
learned correction should also be assessed after propagation through the RANS
equations.

We use a frozen-field propagation approach similar to that adopted for
data-driven turbulence corrections in
\cite{geneva2019quantifying,buchanan2026bayesian}. Independent samples are drawn
from the posterior distribution of the closure, and each sample is evaluated
over the computational mesh to obtain a fixed correction field. The closure is
trained only within the criterion-defined wake region but, as described in
Section~\ref{sec:wakes:setup}, is evaluated over the full domain at deployment.
The RANS solver is then run to convergence once for each fixed correction field.
This differs from the coupled propagation strategy of
\cite{cherroud2022sparse}, in which a stochastic algebraic model is embedded in
the solver and its coefficient uncertainty is propagated using non-intrusive
polynomial-chaos collocation.

During each RANS solve, the sampled correction field is held fixed and is not
recomputed as the mean flow evolves. The ensemble of converged solutions then
provides a distribution of the propagated mean-flow quantities. We refer to
predictions evaluated without the solver as \textit{a priori}, while
\emph{frozen propagation} denotes the present procedure in which one converged
RANS solution is obtained for each fixed sampled correction field. Fully coupled
or online propagation, in which the learned correction is reevaluated during the
nonlinear solver iterations, is not considered here.

\begin{figure}[htbp]
  \centering
  \includegraphics[width=\textwidth]{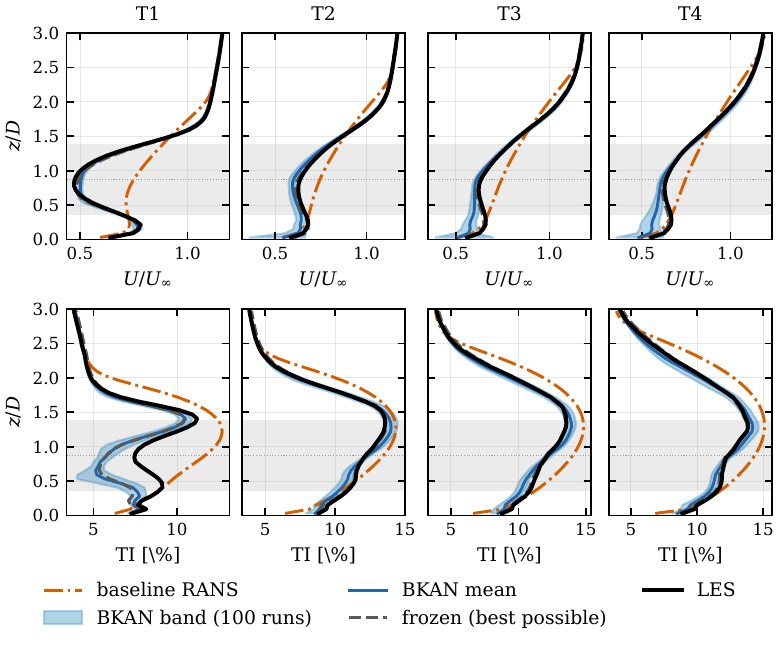}
  \caption{Aligned array (6T5D): vertical profiles at the wake centre, $4D$
  downstream of the first four turbines. Top row, normalised streamwise velocity
  $U/U_\infty$; bottom row, turbulence intensity. Baseline RANS (dash-dot),
  propagated BKAN ensemble (mean and shaded band), frozen best-possible solution
  (dashed), and LES reference (solid); the grey band marks the rotor-swept
  region.}
  \label{fig:prop_6t5d}
\end{figure}

\begin{figure}[htbp]
  \centering
  \includegraphics[width=\textwidth]{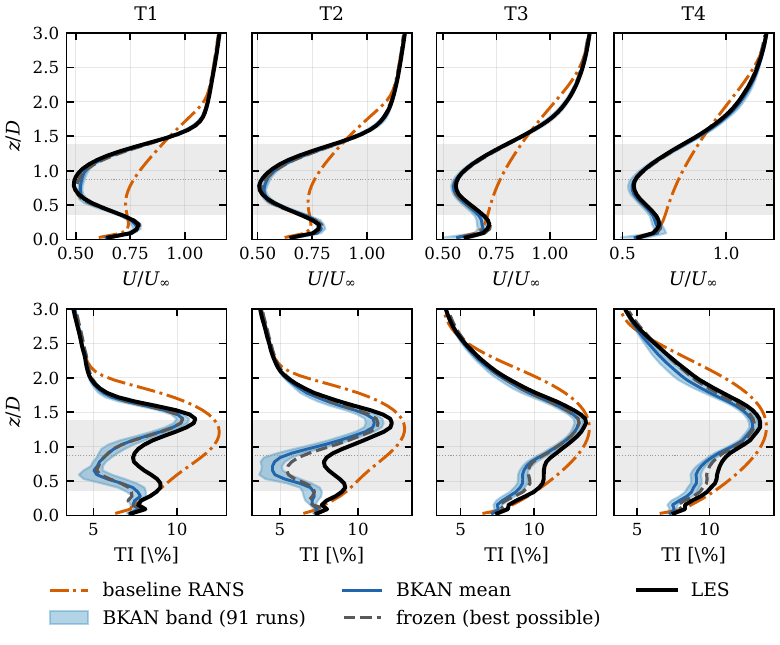}
  \caption{Staggered array (6T7DSTG): vertical profiles at the wake centre, $4D$
  downstream of the first four turbines. Rows and legend as in
  Figure~\ref{fig:prop_6t5d}.}
  \label{fig:prop_stg}
\end{figure}

Figures~\ref{fig:prop_6t5d} and \ref{fig:prop_stg} show vertical profiles of
streamwise velocity (top row) and turbulence intensity (bottom row) at the wake
centre, $4D$ downstream of the first four turbines. The comparisons include the baseline RANS solution, the
propagated BKAN ensemble, a frozen ``best-possible'' solution obtained by
injecting the correction derived from the LES reference, and the LES itself.
The velocity is normalised by the inflow speed,
$U_\infty=8$~m/s at the turbine hub height.

For both unseen layouts, the baseline RANS predicts a weaker wake deficit than
the LES across much of the rotor-swept region. Propagating the BKAN correction
substantially improves the streamwise-velocity prediction. As summarised in
Table~\ref{tab:prop_rmse}, the rotor-region RMSE decreases from $0.119$ to
$0.034$ for the aligned 6T5D array, corresponding to a reduction of approximately
$72\%$, and from $0.170$ to $0.015$ for the staggered 6T7DSTG array, a reduction
of approximately $91\%$. For the staggered case, the BKAN result approaches the
frozen best-possible value of $0.012$. Neither deployment layout is included in
the training set, which contains only the aligned 6T7D array. The propagated
results therefore test transfer to a different turbine spacing and, for
6T7DSTG, to a different arrangement.

The propagated ensemble accounts for both epistemic and aleatoric contributions.
For each realisation, the epistemic component is obtained from one posterior draw
of the Bayesian closure parameters. The variance head is evaluated at the
posterior-mean parameters to obtain the pointwise aleatoric variance
$\sigma_c^2(\bm{\lambda})$ for each stored component $c$ of the symmetric
correction tensor. Before propagation, the corresponding standard deviation
$\sigma_c$ is restricted to the criterion-defined wake region and clipped at its
$99$th percentile within that region. This mask applies only to the aleatoric
perturbation; the BKAN correction itself remains evaluated over the entire
computational domain.

Spatial correlation is imposed on the aleatoric component to represent the
spatial structure of the model discrepancy. For each tensor component, an
independent standard-normal field $w_c(\bm{x})$ is generated over the structured
mesh and passed through an anisotropic Gaussian filter. The resulting field is
recentered and rescaled to unit global variance before being multiplied by the
local predicted standard deviation,
\begin{equation}
  a_c(\bm{x})
  =
  \sigma^{\mathrm{use}}_c(\bm{x})
  \frac{
    \mathcal{G}_{\bm{s}}\!\left[w_c\right](\bm{x})
    -
    \left\langle\mathcal{G}_{\bm{s}}\!\left[w_c\right]\right\rangle
  }{
    \operatorname{std}
    \left(
      \mathcal{G}_{\bm{s}}\!\left[w_c\right]
    \right)
  },
  \qquad
  w_c(\bm{x})\sim\mathcal{N}(0,1),
  \label{eq:correlated-aleatoric}
\end{equation}
where $\mathcal{G}_{\bm{s}}$ denotes anisotropic Gaussian filtering and
$\sigma^{\mathrm{use}}_c$ is zero outside the wake region. Separate random fields
are generated for the six stored tensor components; no cross-component
correlation is imposed. Since a separate field is added per stored component, individual samples are not exactly trace-free; the isotropic part is dynamically inert in the incompressible solver (absorbed into the pressure) and does not affect the propagated statistics.

The spatial correlation scales are estimated once from the \textit{a priori}
residual field of the 6T7D training case. Directional autocorrelations of the
residual are evaluated over the wake region, and the correlation length in each
coordinate direction is defined by the first lag at which the autocorrelation
falls below $e^{-1}$. This gives
\begin{equation}
  (L_x,L_y,L_z)=(600,50,40)\ {\rm m}.
\end{equation}
The same correlation lengths are used for all tensor components. For the
Gaussian filter, they are converted to kernel standard deviations according to
\begin{equation}
  s_j=\frac{L_j}{2\Delta_j},
\end{equation}
where $\Delta_j$ is the grid spacing in direction $j$. The correlation model is
therefore determined entirely from the training case and does not use
high-fidelity information from either deployment layout.

Each propagated realisation consists of one posterior draw of the BKAN
parameters together with one independently generated correlated aleatoric field
for each tensor component. Their sum is written once as a frozen correction
field and held fixed during the corresponding RANS solve. In both deployment
layouts, the LES velocity profile lies within the resulting propagated band over
much of the rotor-swept region
(Figures~\ref{fig:prop_6t5d} and \ref{fig:prop_stg}). The band is of a
magnitude comparable to the observed BKAN--LES discrepancy, although these
profile comparisons should not be interpreted as an independent calibration
test.

\begin{table}[htbp]
  \centering
  \caption{Rotor-region streamwise-velocity root-mean-square error against LES,
  $4D$ downstream of each turbine for the two unseen wind-farm layouts. Results
  are shown for the baseline RANS solution, the propagated BKAN prediction, and
  the frozen best-possible solution. Percentages indicate the reduction relative
  to the baseline RANS error.}
  \label{tab:prop_rmse}
  \begin{tabular}{lccc}
    \toprule
    Layout & Baseline & BKAN & Best-possible \\
    \midrule
    6T5D (aligned, unseen spacing)      & $0.119$ & $0.034$ ($-72\%$) & $0.012$ \\
    6T7DSTG (staggered, unseen layout) & $0.170$ & $0.015$ ($-91\%$) & $0.012$ \\
    \bottomrule
  \end{tabular}
\end{table}

\section{Implications and limitations}
\label{sec:discussion}

The results show that the proposed closure can combine an inspectable
constitutive representation with predictive uncertainty without sacrificing
the accuracy obtained by the corresponding neural baseline. They also reveal
important limits. In particular, the quality of the predicted uncertainty
depends strongly on whether the deployment conditions are represented by the
training data. This section discusses these findings, the choice of Bayesian
parameterisation, and the main limitations of the present formulation.

\subsection{Interpretability and Bayesian parameterisation}
\label{sec:discussion:interpretability}

The comparison with the tensor-basis BNN shows that the interpretability of the
KAN representation is obtained without a loss of predictive accuracy. In
distribution, the BKAN and BNN achieve similar anisotropy errors, while the BKAN
has the lower calibration error among the models compared in
Section~\ref{sec:aniso:indist}. At the same time, the learned univariate edge
functions can be inspected directly and, where their structure is sufficiently
simple, approximated by closed-form expressions. The Bayesian treatment retains
this property by placing distributions over the learned functions rather than
replacing them with an opaque probabilistic mapping.
This combination differs from the sparse Bayesian algebraic approach of
\cite{cherroud2022sparse}. There, uncertainty is introduced into an explicit
closure selected from a prescribed dictionary of candidate terms. In the present
approach, the scalar constitutive functions are instead learned directly through
the KAN representation and can subsequently be inspected or distilled. This
provides greater flexibility in the functional form, at the cost of a more
complex learned model.

The parameterisation study further shows that the location of the posterior
within the KAN matters. Treating the spline coefficients as Bayesian variables
(b-coef) provides the most favourable overall balance of predictive accuracy,
calibration, parameter cost, and interpretability. Placing the posterior only on
the edge amplitudes (b-scale) gives poorer calibration, while treating both the
spline coefficients and edge amplitudes as Bayesian variables (b-full) increases
the epistemic spread and parameter count without improving calibration over
b-coef. Although the detailed ordering of b-scale and b-full varies between
experiments, b-coef is the preferred parameterisation in both the controlled
regression study and the turbulence-anisotropy ablation
(Section~\ref{sec:benchmark}; \ref{app:extra},
Table~\ref{tab:appF:which}). We therefore use b-coef as the default
parameterisation throughout the turbulence applications.

\subsection{Trust boundary under geometric extrapolation}
\label{sec:discussion:trust}

The most important limitation concerns the behaviour of the predictive
uncertainty under extrapolation. In the in-distribution tests considered here,
the uncertainty is well calibrated and provides useful information about the
model error: regions assigned larger uncertainty also tend to contain larger
prediction errors. This behaviour changes when the closure is applied to a
geometry that is absent from the training data. As shown in
Section~\ref{sec:aniso:ood}, the epistemic uncertainty increases in regions
affected by the geometric change, but not enough to account for the accompanying
increase in prediction error. The result is substantial under-coverage and a
marked deterioration in error ranking.

The model therefore retains some sensitivity to extrapolation without providing
a quantitatively reliable measure of its consequences. An increase in epistemic
uncertainty alone should not be interpreted as evidence that the predictive
distribution remains calibrated outside the training regime. For the present
closure, the in-distribution results support the use of the uncertainty estimate
for interpolation within represented flow families, whereas uncertainty estimates
for substantially different geometries should be interpreted cautiously.

Similar under-coverage under extrapolation has been reported for Bayesian
turbulence corrections by \cite{buchanan2026bayesian}, while
\cite{cherroud2022sparse} found that uncertainty intervals can widen in regions
of larger discrepancy without necessarily capturing the full error. One likely
contributor in the present model is the mean-field variational approximation,
which neglects posterior correlations and may underestimate epistemic uncertainty
when the model is taken away from the training distribution.

Several extensions could address this limitation. Posterior tempering can increase
the epistemic spread, although any improvement in coverage must be balanced
against the resulting loss of in-distribution sharpness. Ensembles of independently
trained BKANs could capture variation between different posterior modes that is
not represented by a single mean-field approximation. More expressive
variational families, such as low-rank or full-covariance Gaussian posteriors and
normalising-flow approximations, could instead relax the posterior independence
assumption directly. A complementary approach is to broaden the training set with
flows that contain the geometric and physical variations expected at deployment.
A systematic comparison of these strategies using the same calibration,
sharpness, and sparsification measures employed here would be a useful direction
for future work.

\subsection{Realisability is regularised, not guaranteed}
\label{sec:discussion:realiz}

The eigenvalue penalty in \eqref{eq:realiz} is a soft regularisation term
rather than a hard constraint. The fraction of non-realisable predictions is
already small in distribution without the penalty, and at the training budget
used here the penalty does not materially reduce this fraction or improve the
prediction error (\ref{app:extra}). The tensor-basis representation enforces
invariance, symmetry, and tracelessness, but it does not constrain the
eigenvalues of the predicted anisotropy to the realisable range.

A strict guarantee would therefore require a different parameterisation, for
example one based on barycentric coordinates constrained to the admissible
simplex or on a spectral representation that bounds the eigenvalues by
construction. Such formulations are a natural extension of the present
framework. The signed correction $b_{ij}^{\Delta}$ used for the wind-farm cases
in Section~\ref{sec:wakes} is not itself subject to the same boundedness
requirement.

\subsection{Scope of the present study}
\label{sec:discussion:limitations}

Three aspects define the scope of the present results. First, the solver-level
predictions are obtained using frozen propagation. Each sampled correction field
is held fixed while the RANS equations converge, rather than being recomputed as
the mean flow evolves. The results therefore demonstrate the effect of propagating
the learned posterior through the solver, but do not address fully coupled
\textit{a posteriori} deployment.

Second, the symbolic distillation demonstrated here concerns individual,
low-dimensional KAN functions. It provides a direct way to inspect and
approximate learned constitutive relationships, but does not imply that the
complete tensor-basis closure can always be reduced to a compact analytical
expression. This distinction is particularly relevant where the tensor-basis
coefficients are not uniquely identifiable.

Third, propagation of the aleatoric component requires a spatial correlation
model in addition to the pointwise heteroscedastic variance predicted by the
BKAN. In the present calculations, the correlation length is inferred from the
training-case residual field. The sensitivity of the propagated uncertainty to
this modelling choice has not been examined and should be considered in future
solver-level uncertainty studies.

\section{Concluding remarks}
\label{sec:conclusions}

We have developed a Bayesian tensor-basis KAN for Reynolds-stress modelling that
combines an inspectable constitutive representation with predictive uncertainty.
Learnable spline functions on the KAN edges provide direct access to the scalar
relationships underlying the closure and, where their structure is sufficiently
simple, allow \textit{post hoc} symbolic approximation. The variational treatment
provides epistemic and aleatoric components of predictive uncertainty, while the
tensor-basis formulation enforces Galilean and rotational invariance, symmetry,
and tracelessness by construction.

The model was evaluated progressively from controlled regression to turbulent
flows. On the regression benchmark, the BKAN achieves predictive performance
comparable to the Bayesian neural baseline while providing well-calibrated
uncertainty. Among the Bayesian KAN parameterisations considered, placing the
posterior on the spline coefficients (b-coef) gives the most favourable overall
balance of predictive accuracy, calibration, parameter cost, and
interpretability. For Reynolds-stress anisotropy, the closure is accurate in
distribution and its uncertainty is both well calibrated and informative: larger
predicted uncertainty tends to coincide with larger prediction errors. This
behaviour deteriorates under geometric extrapolation. Although the epistemic
uncertainty responds to unfamiliar flow regions, it does not increase sufficiently
to account for the corresponding prediction error, leading to substantial
under-coverage and poorer error ranking. This defines a practical trust boundary
for the present uncertainty model.

The three-dimensional wind-farm cases demonstrate that the same framework can be
applied to an anisotropy correction and propagated through a RANS solver.
Corrections learned from a six-turbine array transfer successfully to unseen
wind-farm layouts, and frozen propagation substantially improves the resulting
wake-velocity predictions. Relative to the baseline RANS solution, the
rotor-region streamwise-velocity RMSE is reduced by approximately $72\%$ for the
unseen aligned 6T5D array and $91\%$ for the unseen staggered 6T7DSTG array.
These results show that the posterior can be propagated beyond the closure itself
to quantify uncertainty in flow quantities of practical interest.

The main remaining challenges follow directly from the limitations identified in
this study. Future work should focus on posterior approximations that better
represent epistemic uncertainty under geometric shift, and fully coupled solver propagation in which the learned
correction is updated as the flow evolves. These extensions would broaden the
range over which the uncertainty estimate can be used reliably while retaining
the interpretability provided by the KAN representation.

\section*{Acknowledgments}
A.E. acknowledges support from the NWO Open Competition ENW-XS project U-TUBKAN (project number OCENW.XS25.3.099), funded by the Dutch Research Council (NWO).

\section*{CRediT authorship contribution statement}
A. E.: Conceptualization (lead); Methodology (lead); Software \& code (lead); Formal analysis (lead); Writing – original draft (lead).
Kh. J.: Conceptualization (supporting); Methodology (supporting); Software \& code (supporting); Formal analysis (equal); Writing – original draft (supporting).
T. B.: Conceptualization (supporting); Methodology (supporting); Software \& code (supporting); Formal analysis (supporting); Writing – original draft (supporting).
R. P. D.: Conceptualization (equal); Methodology (equal); Supervision (lead); Writing – review and editing (equal).

\section*{Code and data availability}
A reference implementation of the Bayesian Kolmogorov--Arnold network, together with the regression-benchmark and \textit{a priori} anisotropy experiments, is openly available at \url{https://github.com/alieidi1st/bayesian-kan-turbulence} under the MIT license. The UCI regression datasets follow the standard splits of \cite{hernandez2015probabilistic}, and the turbulence-anisotropy study uses the curated dataset of \cite{mcconkey2021curated}; both are obtained from their original public sources as documented in the repository. The wind-farm large-eddy-simulation cases are not redistributed owing to their size and are available from the authors on reasonable request.

\section*{Declaration of competing interest}
The authors declare that they have no known competing financial interests or personal relationships that could have appeared to influence the work reported in this paper.

\appendix
\makeatletter
\@addtoreset{figure}{section}
\@addtoreset{table}{section}
\makeatother
\renewcommand{\thefigure}{\thesection\arabic{figure}}
\renewcommand{\thetable}{\thesection\arabic{table}}
\section{Model formulation and training}
\label{app:model}

\subsection{Variational formulation}
\label{app:vi}

We place a mean-field Gaussian posterior
\begin{equation}
  q_\psi(\bm{\theta})
  =
  \prod_i
  \mathcal{N}(\theta_i\mid\mu_i,\sigma_i^2)
\end{equation}
over the parameters treated as Bayesian variables, as described in
Section~\ref{sec:method:which}. The standard deviation is parameterised as
$\sigma_i=\mathrm{softplus}(\rho_i)$ to ensure positivity. The model is trained
by maximising the ELBO in \eqref{eq:elbo}.

The expected log-likelihood is estimated by Monte Carlo sampling. For the
heteroscedastic Gaussian likelihood over the $d$ unique anisotropy components,
the negative expected log-likelihood is approximated by
\begin{equation}
  -\E_q\!\left[\log p(\mathcal{D}\mid\bm{\theta})\right]
  \approx
  \frac{1}{2}
  \sum_{\mathrm{pts}}
  \sum_{c=1}^{d}
  \left[
    \log\sigma_c^2(\bm{\lambda})
    +
    \frac{\big(b_c-\hat b_c\big)^2}
         {\sigma_c^2(\bm{\lambda})}
  \right]
  + \mathrm{const}.
\end{equation}
For the Gaussian prior $p(\theta_i)=\mathcal{N}(0,s^2)$, the KL term has the
closed-form expression
\begin{equation}
  \KL(q_\psi\|p)
  =
  \sum_i
  \left[
    \log\frac{s}{\sigma_i}
    +
    \frac{\sigma_i^2+\mu_i^2}{2s^2}
    -
    \frac12
  \right].
\end{equation}

To reduce gradient variance, the local reparameterisation trick is applied to
the spline output of each edge. For the B-spline branch of the edge in
\eqref{eq:edge}, with Gaussian coefficients
$\theta_k\sim\mathcal{N}(\mu_k,\sigma_k^2)$, the spline output at input $x$ is
itself Gaussian,
\begin{equation}
  \phi_s(x)
  \sim
  \mathcal{N}
  \left(
    \sum_k B_k(x)\mu_k,\;
    \sum_k B_k(x)^2\sigma_k^2
  \right),
  \label{eq:localreparam}
\end{equation}
so the scalar spline output can be sampled directly rather than sampling every
coefficient $\theta_k$ individually. This reduces the variance of the gradient
estimator during variational training. Edge-amplitude posteriors in the b-scale
and b-full variants are sampled directly.

For the two-component scale-mixture prior
\begin{equation}
  p(\theta)
  =
  \pi\,\mathcal{N}(0,s_1^2)
  +
  (1-\pi)\,\mathcal{N}(0,s_2^2),
\end{equation}
the KL term has no closed form and is estimated using the parameters sampled
during the forward pass,
\begin{equation}
  \KL
  \approx
  \sum_i
  \left[
    \log q_\psi(\theta_i)-\log p(\theta_i)
  \right].
\end{equation}
The Gamma hyperprior
$p(s^{-2})=\mathrm{Gamma}(a_0,b_0)$ gives, through the evidence-maximising
update of \cite{mackay1992practical}, the adaptive prior precision
\begin{equation}
  \lambda^\star
  =
  \frac{a_0+n/2}
       {b_0+\tfrac12\,\E_q\!\left[\sum_i\theta_i^2\right]},
  \qquad
  s^2=1/\lambda^\star,
  \label{eq:ard}
\end{equation}
where $n$ is the number of parameters in the corresponding group.
$\lambda^\star$ is recomputed during training and substituted into the
Gaussian KL term. In all cases, the KL contribution is scaled by $1/N$ and
annealed from zero to one during training, as in \eqref{eq:loss}. We found
no material sensitivity of the results to the three prior formulations and
therefore use the Gaussian prior unless stated otherwise.

At prediction time, we draw $S$ posterior samples from $q_\psi$, reconstruct
the anisotropy through the fixed tensor-basis merge in \eqref{eq:tb} for
each sample, and estimate the epistemic variance from the resulting ensemble.
This is combined with the mean predicted aleatoric variance according to the
law of total variance in \eqref{eq:totvar}. The reported point prediction
is evaluated at the posterior-mean parameters. The complete training and
prediction procedure is summarised in Algorithm~\ref{alg:train}.

\subsection{Tensor basis, invariants and features}
\label{app:basis}

Let $\widehat{\bm{S}}$ and $\widehat{\bm{W}}$ denote the mean strain- and
rotation-rate tensors normalised by a turbulent time scale. The integrity basis
of \cite{pope1975more} defines a set of tensor basis elements
$\bm{T}^{(n)}$ and scalar invariants $\lambda_p$. For two-dimensional mean
flows, the following three-term basis is exact:
\begin{equation}
  \left\{
    \widehat{\bm{S}},\quad
    \widehat{\bm{S}}\widehat{\bm{W}}
      -\widehat{\bm{W}}\widehat{\bm{S}},\quad
    \widehat{\bm{S}}^2
      -\tfrac13\mathrm{tr}(\widehat{\bm{S}}^2)\bm{I}
  \right\},
\end{equation}
with invariants
\begin{equation}
  \lambda_1=\mathrm{tr}(\widehat{\bm{S}}^2),
  \qquad
  \lambda_2=\mathrm{tr}(\widehat{\bm{W}}^2).
\end{equation}
The normalised tensors are
\begin{equation}
  \widehat{\bm S}
  =
  \tfrac12\tau
  \left(
    \nabla\bm U+\nabla\bm U^{\!\top}
  \right),
  \qquad
  \widehat{\bm W}
  =
  \tfrac12\tau
  \left(
    \nabla\bm U-\nabla\bm U^{\!\top}
  \right),
\end{equation}
where $\tau$ is the turbulent time scale, taken as $1/\omega$ for
$k$--$\omega$ models and $k/\varepsilon$ for $k$--$\varepsilon$ models.

For three-dimensional flows, including the square duct and the turbine wakes,
we use the full ten-term basis with five scalar invariants,
\begin{equation}
  \lambda_1=\mathrm{tr}(\widehat{\bm S}^2),\qquad
  \lambda_2=\mathrm{tr}(\widehat{\bm W}^2),\qquad
  \lambda_3=\mathrm{tr}(\widehat{\bm S}^3),
\end{equation}
\begin{equation}
  \lambda_4=\mathrm{tr}(\widehat{\bm W}^2\widehat{\bm S}),
  \qquad
  \lambda_5=\mathrm{tr}(\widehat{\bm W}^2\widehat{\bm S}^2).
\end{equation}
The first six basis tensors are
\begin{align}
  \bm T^{(1)} &=
  \widehat{\bm S},
  \\
  \bm T^{(2)} &=
  \widehat{\bm S}\widehat{\bm W}
  -
  \widehat{\bm W}\widehat{\bm S},
  \\
  \bm T^{(3)} &=
  \widehat{\bm S}^2
  -
  \tfrac13\mathrm{tr}(\widehat{\bm S}^2)\bm I,
  \\
  \bm T^{(4)} &=
  \widehat{\bm W}^2
  -
  \tfrac13\mathrm{tr}(\widehat{\bm W}^2)\bm I,
  \\
  \bm T^{(5)} &=
  \widehat{\bm W}\widehat{\bm S}^2
  -
  \widehat{\bm S}^2\widehat{\bm W},
  \\
  \bm T^{(6)} &=
  \widehat{\bm W}^2\widehat{\bm S}
  +
  \widehat{\bm S}\widehat{\bm W}^2
  -
  \tfrac23
  \mathrm{tr}(\widehat{\bm S}\widehat{\bm W}^2)\bm I.
\end{align}
The tensors $\bm T^{(7)}$--$\bm T^{(10)}$ are the remaining higher-order
terms of the integrity basis given by \cite{pope1975more}. Each basis tensor is
unit-normalised before the merge step to improve numerical conditioning.

In addition to the tensor invariants, we use three bounded,
Galilean-invariant scalar markers following
\cite{wang2017comprehensive,wu2018physics}: a strain--rotation balance $q$, a
turbulence-intensity measure $\mathrm{TI}$, and a streamline pressure-gradient
marker $\mathrm{PG}$,
\begin{equation}
  q
  =
  \frac{
    \|\widehat{\bm W}\|^2-\|\widehat{\bm S}\|^2
  }{
    \|\widehat{\bm W}\|^2+\|\widehat{\bm S}\|^2
  },
  \qquad
  \mathrm{TI}
  =
  \frac{k}
       {\tfrac12 U_iU_i+k},
\end{equation}
\begin{equation}
  \mathrm{PG}
  =
  \frac{
    U_k\,\partial_k p
  }{
    \sqrt{U_iU_i\,\partial_jp\,\partial_jp}
    +
    |U_k\,\partial_kp|
  }.
\end{equation}
The ratio forms make these quantities non-dimensional and bounded. They provide
additional information about the local flow state, including turbulence level,
the balance between strain and rotation, and acceleration or deceleration along
the streamline. Together with the tensor invariants, these markers form the
complete feature vector $\bm{\lambda}$ supplied to the BKAN. All scalar network
inputs are standardised to zero mean and unit variance before training.

\subsection{Architectures, training and reproducibility}
\label{app:train}

The closure maps the invariant features to the tensor-basis coefficients using a
KAN. Unless stated otherwise, the KAN has one hidden layer
of width sixteen and uses cubic B-splines on a five-knot grid. A second KAN head,
with the same architecture, predicts the heteroscedastic log-variance.

Training consists of two phases. First, the posterior means are deterministically
pre-trained using a mean-squared-error objective. This short pre-training stage is
needed for the network to resolve sharp fields. Second, the model is trained
variationally by minimising the loss in \eqref{eq:loss}. We use Adam optimizer with batch
size $8192$, learning rate $5\times10^{-3}$, and linear KL annealing.

\begin{figure}[htbp]
  \centering
  \includegraphics[width=0.82\textwidth]{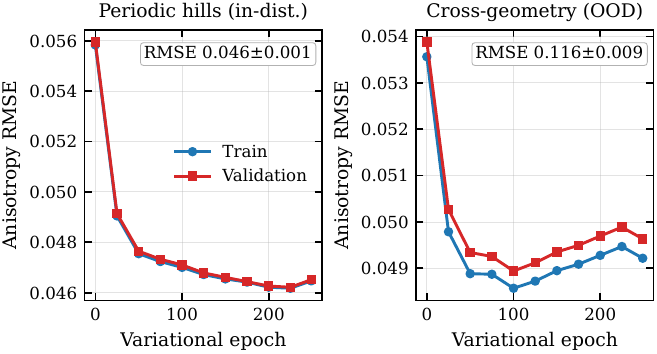}
  \caption{Train and validation RMSE convergence for seed~0 in the in-distribution
  and out-of-distribution settings. In distribution, the train and validation
  curves remain close, indicating no evident over-fitting. The persistent
  out-of-distribution gap reflects the difficulty of geometric extrapolation.}
  \label{fig:appA:conv}
\end{figure}

In-distribution metrics are averaged over three random seeds. A $15\%$ validation
hold-out is used to monitor over-fitting, and the corresponding train and
validation convergence is shown in Figure~\ref{fig:appA:conv}. Predictive moments
are computed using $S=50$--$100$ posterior samples, following
\eqref{eq:totvar}. Reported calibration metrics exclude the cross-plane shear
components of planar flows, which are degenerate and identically close to zero.
Including these components would artificially inflate the calibration metric.

The complete set of hyperparameters for each experiment is given in
Table~\ref{tab:hyper}. The regression-benchmark width is matched to the reference
BNN as described in \ref{app:uci}; the turbulence and turbine
models share a single architecture, with the realisability penalty active only on
the bounded anisotropy target and disabled for the signed wake correction. No
hyperparameter was tuned on the out-of-distribution or test splits; the validation
hold-out was used only to confirm the absence of over-fitting.

\begin{table}[htbp]
  \centering
  \caption{Hyperparameters and training protocol by experiment. ``Det.\ / VI
  epochs'' are the deterministic pre-training and variational-inference epoch
  counts; $\alpha$ is the realisability-penalty weight; $S$ is the number of
  posterior samples used for prediction.}
  \label{tab:hyper}
  \small
  \begin{tabular}{lcccc}
    \toprule
    Setting & UCI benchmark & Planar anisotropy & Square duct & Turbine wakes \\
    \midrule
    Hidden layer(s)        & matched width & $[16]$ & $[16]$ & $[16]$ \\
    Spline order / knots   & cubic / 5 & cubic / 5 & cubic / 5 & cubic / 5 \\
    Tensor-basis terms     & --- & 3 & 10 & 10 \\
    Invariants + markers    & dataset inputs & $2+3$ & $5+3$ & $5+3$ \\
    Noise model            & homoscedastic & heteroscedastic & heteroscedastic & heteroscedastic \\
    Bayesian variant       & b-coef & b-coef & b-coef & b-coef \\
    Prior                  & Gaussian & Gaussian & Gaussian & Gaussian \\
    Det.\ / VI epochs      & --- / $400$\,$^\dagger$ & $60$ / $250$ & $60$ / $250$ & $60$ / $250$ \\
    Batch size             & $128$ & $8192$ & $8192$ & $8192$ \\
    Learning rate          & $10^{-2}$ & $5\times10^{-3}$ & $5\times10^{-3}$ & $5\times10^{-3}$ \\
    KL anneal (epochs)     & $E/4$ & $100$ & $100$ & $100$ \\
    Realisability $\alpha$ & $0$ & $10$ & $10$ & $0$ \\
    Predictive samples $S$ & $100$ & $50$--$100$ & $50$--$100$ & $100$ \\
    Optimiser              & Adam & Adam & Adam & Adam \\
    \bottomrule
  \end{tabular}\\[2pt]
  {\footnotesize $^\dagger$ per-dataset schedule; see \ref{app:uci}.}
\end{table}

\section{Datasets and flow cases}
\label{app:data}

\subsection{Regression benchmark}
\label{app:uci}

The regression benchmark follows \cite{hernandez2015probabilistic}. We use the
standard datasets listed in Table~\ref{tab:uci}, apply $z$-score
standardisation, and use the standard random train--test splits. Errors and
log-likelihoods are reported in the original target units.

For each dataset, the KAN width is chosen as the largest single-hidden-layer
width for which the number of trainable parameters does not exceed that of the
reference one-hidden-layer BNN. The comparison is therefore performed at a
matched parameter budget. The Bayesian-parameterisation ablation shown in
Figure~\ref{fig:whichparams} uses five of the benchmark datasets, with five
random splits for each dataset.

\begin{table}[htbp]
  \centering
  \caption{Regression benchmark datasets following
  \cite{hernandez2015probabilistic}: input dimension $d$, number of instances
  $N$, and number of standard splits.}
  \label{tab:uci}
  \begin{tabular}{lccc}
    \toprule
    Dataset & $d$ & $N$ & Splits \\
    \midrule
    Boston housing            & 13 & 506    & 20 \\
    Concrete                  &  8 & 1030   & 20 \\
    Energy                    &  8 & 768    & 20 \\
    Kin8nm                    &  8 & 8192   & 20 \\
    Naval propulsion          & 16 & 11934  & 20 \\
    Power plant               &  4 & 9568   & 20 \\
    Wine (red)                & 11 & 1599   & 20 \\
    Yacht                     &  6 & 308    & 20 \\
    Protein                   &  9 & 45730  & 5  \\
    \bottomrule
  \end{tabular}
\end{table}

Table~\ref{tab:uci:rmse} and Table~\ref{tab:uci:ll} report the per-dataset test
RMSE and test log-likelihood, averaged over the standard splits. The BKAN is
competitive with the other models on both metrics across the suite; the
corresponding calibration summary is shown in Figure~\ref{fig:uci}.

\begin{table}[htbp]
  \centering
  \caption{Per-dataset test RMSE (mean over the standard splits) on the
  regression benchmark; lower is better.}
  \label{tab:uci:rmse}
  \small
  \begin{tabular}{lcccccc}
    \toprule
    Dataset & BKAN & BNN & MLP & KAN & MC-drop. & Deep ens. \\
    \midrule
    Boston   & 3.367 & 3.159 & 3.504 & 4.576 & 3.134 & 3.144 \\
    Concrete & 4.968 & 5.177 & 5.195 & 5.527 & 5.190 & 4.824 \\
    Energy   & 0.378 & 0.498 & 0.440 & 0.341 & 1.355 & 0.403 \\
    Kin8nm   & 0.088 & 0.075 & 0.079 & 0.086 & 0.096 & 0.074 \\
    Naval    & 0.000 & 0.001 & 0.001 & 0.000 & 0.001 & 0.000 \\
    Power    & 3.894 & 4.015 & 3.973 & 3.905 & 4.109 & 3.945 \\
    Wine     & 0.674 & 0.633 & 0.679 & 0.847 & 0.642 & 0.632 \\
    Yacht    & 0.551 & 0.801 & 0.577 & 0.551 & 0.697 & 0.480 \\
    Protein  & 4.124 & 4.402 & 4.472 & 4.118 & 4.585 & 4.405 \\
    \bottomrule
  \end{tabular}
\end{table}

\begin{table}[htbp]
  \centering
  \caption{Per-dataset test log-likelihood (mean over the standard splits) for
  the probabilistic models; higher is better.}
  \label{tab:uci:ll}
  \small
  \begin{tabular}{lcccc}
    \toprule
    Dataset & BKAN & BNN & MC-drop. & Deep ens. \\
    \midrule
    Boston   & $-2.93$ & $-2.56$ & $-2.51$ & $-3.93$ \\
    Concrete & $-3.15$ & $-3.07$ & $-2.97$ & $-2.91$ \\
    Energy   & $-0.47$ & $-0.77$ & $-1.90$ & $-1.04$ \\
    Kin8nm   & $ 1.01$ & $ 1.17$ & $ 0.99$ & $ 1.30$ \\
    Naval    & $ 6.87$ & $ 5.83$ & $ 4.75$ & $ 7.08$ \\
    Power    & $-2.78$ & $-2.81$ & $-2.82$ & $-2.76$ \\
    Wine     & $-1.17$ & $-0.96$ & $-2.09$ & $-1.35$ \\
    Yacht    & $-0.89$ & $-1.24$ & $-0.31$ & $ 0.55$ \\
    Protein  & $-2.84$ & $-2.90$ & $-2.83$ & $-2.77$ \\
    \bottomrule
  \end{tabular}
\end{table}

\subsection{Turbulence-anisotropy cases}
\label{app:turbcases}

The Reynolds-stress anisotropy study in Section~\ref{sec:anisotropy} uses the
curated dataset of \cite{mcconkey2021curated}, which provides matched
baseline-RANS ($k$--$\omega$ SST) and high-fidelity (DNS/LES) fields interpolated
onto common grids for a family of canonical flows. The cases used here are
summarised in Table~\ref{tab:turbcases}. The planar flows use the two-invariant
three-term basis and the square duct the five-invariant ten-term basis.

In-distribution results use the parametric periodic-hill family, in which the
hill slope is rescaled by a parameter $\alpha$: the model is trained on
$\alpha\in\{0.5,0.8,1.2,1.5\}$ and tested on the held-out $\alpha=1.0$ geometry
at the same bulk Reynolds number. The out-of-distribution test instead withholds
entire geometries: the model is trained on the periodic-hill and
zero-pressure-gradient flat-plate flows and evaluated on the
converging--diverging channel and the curved backward-facing step, neither of
which appears in training. The square duct is treated separately, with training
over a range of bulk Reynolds numbers and a held-out Reynolds number, to test the
three-dimensional secondary-flow anisotropy that a linear eddy-viscosity model
cannot represent. All in-distribution anisotropy metrics are reported as means
over three random seeds.

\begin{table}[htbp]
  \centering
  \caption{Turbulence-anisotropy cases from the curated dataset of
  \cite{mcconkey2021curated}. ``In-dist.'' denotes interpolation within a flow
  family and ``OOD'' a withheld geometry.}
  \label{tab:turbcases}
  \small
  \begin{tabular}{lll}
    \toprule
    Case & Configuration & Role \\
    \midrule
    Periodic hills ($\alpha{=}0.5,0.8,1.2,1.5$) & parametric slope, $Re_b{=}5600$, 2-D & in-dist.\ train \\
    Periodic hills ($\alpha{=}1.0$)             & baseline slope, $Re_b{=}5600$, 2-D   & in-dist.\ test \\
    Flat plate (ZPG boundary layer)             & $Re_\theta{\approx}1000$--$4000$, 2-D & OOD train pool \\
    Converging--diverging channel               & $Re{\approx}12\,600,\,20\,580$, 2-D   & OOD test \\
    Curved backward-facing step                 & $Re{\approx}13\,700$, 2-D             & OOD test \\
    Square duct                                 & $Re_b{=}1100$--$3500$, 3-D            & secondary flow \\
    \bottomrule
  \end{tabular}
\end{table}

\subsection{Wind-farm cases and wake-region criterion}
\label{app:turbine}

The wind-farm dataset contains RANS solutions and matched high-fidelity
references for the four cases summarised in Table~\ref{tab:turbine}. The
training samples for the anisotropy correction are restricted to the wake
region, while at deployment the trained closure is evaluated over the full
computational domain, as described in Section~\ref{sec:wakes:setup}.

Following \cite{jigjid2025discovery}, the wake region is identified using a
threshold on the magnitude of the induced correction to turbulent-kinetic-energy
production,
\begin{equation}
  |\mathcal{P}^{\Delta}|>\tau_P,
  \qquad
  \mathcal{P}^{\Delta}
  =
  -2k\,b^{\Delta}_{ij}\,\partial_j U_i,
  \label{eq:app:wakecriterion}
\end{equation}
where $\tau_P$ denotes the prescribed production-correction threshold.

\begin{figure}[htbp]
  \centering
  \includegraphics[width=0.78\textwidth]{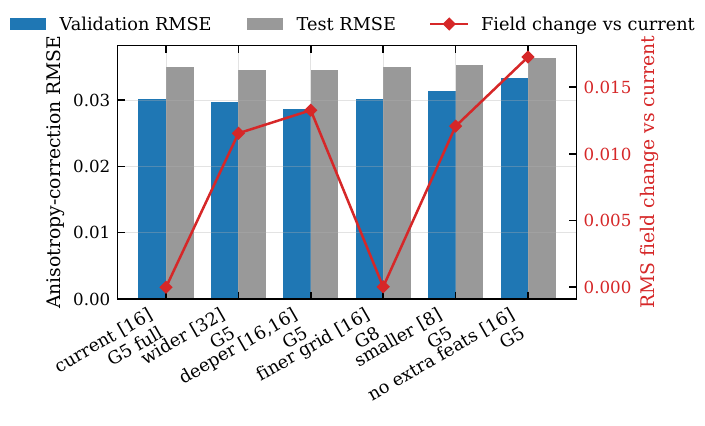}
  \caption{Architecture sensitivity for the wind-farm correction model.
  Validation and test RMSE vary little with network width, depth, and
  spline-grid resolution. The RMS change in the deployed correction field is
  also small, with the largest change occurring when the additional scalar
  features are removed.}
  \label{fig:appE:sens}
\end{figure}

Figure~\ref{fig:appE:sens} reports the architecture-sensitivity study used to
select the configuration adopted for the wind-farm calculations. Variations in
network width, depth, and spline-grid resolution produce only small changes in
the validation and test RMSE. The RMS change in the deployed correction field
is also small relative to the selected configuration. Removing the additional
scalar features has the largest effect among the variants considered, giving
both the highest validation and test errors and the largest change in the
predicted correction field, although the absolute change remains modest.

\begin{table}[htbp]
  \centering
  \caption{Wind-farm cases. Each case is a full three-dimensional
  atmospheric-boundary-layer RANS solution with a matched high-fidelity
  reference.}
  \label{tab:turbine}
  \begin{tabular}{lcc}
    \toprule
    Case & Configuration & Mesh cells \\
    \midrule
    1T       & single turbine                 & $2.73\times10^6$ \\
    6T5D     & six turbines, aligned, $5D$    & $2.73\times10^6$ \\
    6T7D     & six turbines, aligned, $7D$    & $2.73\times10^6$ \\
    6T7DSTG  & six turbines, staggered, $7D$  & $2.73\times10^6$ \\
    \bottomrule
  \end{tabular}
\end{table}

\section{Extended results}
\label{app:extra}

\subsection{Multi-seed results and Bayesian parameterisation}

Table~\ref{tab:appF:seed} reports the three-seed turbulence-anisotropy results
for the b-coef parameterisation, while Table~\ref{tab:appF:which} compares the
three Bayesian parameterisations on the periodic-hill in-distribution split.
The multi-seed results confirm the contrast observed in the main text between
the in-distribution and cross-geometry cases: both the prediction error and the
calibration error increase substantially under geometric extrapolation.

For the parameterisation study, b-coef gives the lowest RMSE and calibration
error among the three variants. The b-scale and b-full alternatives both produce
larger epistemic uncertainty and less favourable predictive performance, although
their detailed ordering differs. The mean epistemic uncertainty increases in the
order
$\text{coef}<\text{scale}<\text{full}$.
Together with the regression benchmark in Section~\ref{sec:benchmark}, these
results support the use of b-coef as the default Bayesian parameterisation.

\begin{table}[htbp]
  \centering
  \caption{Three-seed turbulence-anisotropy results reported as mean $\pm$
  standard deviation.}
  \label{tab:appF:seed}
  \begin{tabular}{lccc}
    \toprule
    Split & Regime & RMSE & Calibration error \\
    \midrule
    Periodic hills  & in-distribution     & $0.046\pm0.001$ & $0.061\pm0.013$ \\
    Cross-geometry  & out-of-distribution & $0.116\pm0.009$ & $0.255\pm0.014$ \\
    \bottomrule
  \end{tabular}
\end{table}

\begin{table}[htbp]
  \centering
  \caption{Bayesian-parameterisation ablation for turbulence anisotropy on the
  periodic-hill in-distribution split. The posterior is placed on the spline
  coefficients (b-coef), the edge amplitudes (b-scale), or both (b-full).
  Among the three variants, b-coef gives the lowest RMSE and calibration error,
  while the mean epistemic uncertainty increases in the order
  coef\,$<$\,scale\,$<$\,full.}
  \label{tab:appF:which}
  \begin{tabular}{lcccc}
    \toprule
    Variant & Parameters & RMSE & Calibration error & Mean epistemic \\
    \midrule
    b-coef  & 5\,472 & $0.045$ & $0.067$ & $0.0019$ \\
    b-scale & 3\,648 & $0.051$ & $0.108$ & $0.0080$ \\
    b-full  & 6\,080 & $0.049$ & $0.122$ & $0.0085$ \\
    \bottomrule
  \end{tabular}
\end{table}

\subsection{Realisability penalty}

Table~\ref{tab:appF:realiz} reports the realisability ablation underlying
Figure~\ref{fig:realiz}. The fraction of non-realisable predictions is already
small in distribution without the penalty. At the training budget considered
here, introducing the soft penalty does not reduce this fraction and has little
effect on the prediction error. Out of distribution, the fraction of
non-realisable states increases to several per cent and is likewise not improved
by the penalty.

The tensor-basis representation enforces invariance, symmetry, and tracelessness,
but does not guarantee that the predicted eigenvalues remain within the
realisable range. These results therefore support the use of a hard
parameterisation if strict realisability is required, as discussed in
Section~\ref{sec:discussion}.

\begin{table}[htbp]
  \centering
  \caption{Non-realisable fraction and test RMSE with the realisability penalty
  off ($\alpha=0$) and on ($\alpha=10$).}
  \label{tab:appF:realiz}
  \begin{tabular}{lcccc}
    \toprule
    & \multicolumn{2}{c}{In-distribution}
    & \multicolumn{2}{c}{Out-of-distribution} \\
    \cmidrule(lr){2-3}\cmidrule(lr){4-5}
    & Non-realisable & RMSE & Non-realisable & RMSE \\
    \midrule
    $\alpha=0$  & $0.09\%$ & $0.0453$ & $4.35\%$ & $0.1077$ \\
    $\alpha=10$ & $0.58\%$ & $0.0451$ & $5.08\%$ & $0.1102$ \\
    \bottomrule
  \end{tabular}
\end{table}

\section*{Declaration of generative AI and AI-assisted technologies in the manuscript preparation process}

During the preparation of this work, the authors used ChatGPT (OpenAI) and Claude (Anthropic) for language editing, wording refinement, and code syntax checking. The authors reviewed and edited the output as needed and take full responsibility for the content of the published article.


\bibliographystyle{elsarticle-num-names}
\bibliography{bibliography,MR}

\end{document}